\documentclass[11pt]{article}
\usepackage{geometry}
\usepackage{graphicx}
\usepackage{amssymb}
\usepackage{epstopdf}
\usepackage{caption}
\usepackage{hyperref}
\usepackage{multicol}
\usepackage{natbib}
\usepackage{booktabs}
\usepackage{amsthm}
\usepackage[english]{babel}
\usepackage{amssymb}
\usepackage{float}
\usepackage{bbm}
\usepackage{empheq}
\usepackage{amsmath}
\usepackage{lipsum}
\usepackage{docmute}
\usepackage{multirow}
\usepackage{setspace} 
\usepackage{hyperref}
\usepackage{subcaption}
\usepackage{mathtools}

\setcitestyle{authoryear,open={(},close={)},citesep={;}}

\providecommand{\keywords}[1]
{
  \small	
  \textbf{\textit{Keywords---}} #1
}

\title{Manifold-Constrained Gaussian Processes for Inference of Mixed-effects Ordinary Differential Equations with Application to Pharmacokinetics}

\author{Yuxuan Zhao and Samuel W.K. Wong\footnote{E-mail for correspondence: samuel.wong@uwaterloo.ca} \\
Department of Statistics and Actuarial Science, University of Waterloo, \\Waterloo, N2L 3G1, Canada.}

\begin{document}
\setstretch{1.25}

\maketitle

\begin{abstract}   
Pharmacokinetic modeling using ordinary differential equations (ODEs) has an important role in dose optimization studies, where dosing must balance sustained therapeutic efficacy with the risk of adverse side effects. 
Such ODE models characterize drug plasma concentration over time and allow pharmacokinetic parameters to be inferred, such as drug absorption and elimination rates. 
For time-course studies involving treatment groups with multiple subjects, mixed-effects ODE models are commonly used. 
However, existing methods tend to lack uncertainty quantification on a subject-level, for key measures such as peak or trough concentration and for making predictions of drug concentration.
To address such limitations, we propose an extension of manifold-constrained Gaussian processes for inference of general mixed-effects ODE models within a Bayesian statistical framework. 
We evaluate our method on simulated examples, demonstrating its ability to provide fast and accurate inference for parameters and trajectories using nested optimization. 
To illustrate the practical efficacy of the proposed method, we provide a real data analysis of a pharmacokinetic model used for an HIV combination therapy study.
\end{abstract}


\section{Introduction}
In clinical drug development, pharmacokinetic studies play a crucial role in optimizing treatment regimens for long-term diseases.
These studies may assign subjects to different treatment groups with varying dosage levels and collect time-course plasma concentration data at discrete time points \citep{wasmuth2004comparison, white2013pharmacokinetic}. An important focus is to quantify key measures of drug exposure within the body based on the observed data, which in turn inform dosing strategies that balance drug safety and efficacy. 
Specifically, higher peak plasma concentration may be associated with an increased risk of severe side effects, while insufficient trough plasma concentration may fail to maintain drug efficacy. Thus, pharmacokinetic modeling can help guide dose optimization across a wide range of therapeutic areas \citep{sy2016pharmacokinetics,lea2018clinical}.

The plasma concentration of the $j$-th subject at time $t$, which we denote by $C_j(t)$, is often modeled as a dynamic system governed by ordinary differential equations (ODEs). 
In pharmacokinetics, the well-known Bateman equation \citep{bateman1983clinical} 
refers to a general framework for constructing differential equations that describe the sequential phases of drug absorption and elimination.
The simplest forms of the Bateman equation only consider drug elimination, 
e.g., the one-compartment model $C_j'(t) = -Ke_j\cdot C_j(t)$, with $Ke_j$ denoting the subject-specific drug elimination rate \citep{sopasakis2018fractional}, along with the standard two-compartment pharmacokinetic model \citep{Talevi2022}. However,
treating the initial drug absorption phase as instantaneous may not always be appropriate, e.g., for oral intake \citep{savic2007implementation}. 
Instead, the general two-compartment version of the Bateman equation can be more widely applicable as it accounts for an absorption delay, which leads to the following ODE \citep{wang2014estimating}:
\begin{equation}\frac{\mathrm{d} C_j(t)}{\mathrm{d} t}=-K e_j C_j(t)+\frac{D_j K e_j K a_j}{C l_j} \exp \left(-K a_j t\right),
    \label{eqn:pk/pd_model}
\end{equation}
where for the $j$-th subject, $D_j$ represents the cumulative amount of unabsorbed drug at the initial time ($t=0$), $Cl_j$ denotes the rate of the total body drug clearance and $Ka_j$ denotes the drug absorption rate.  Moreover, parameters (such as $Ka_j$ and $Cl_j$) may exhibit substantial between-subject variation within the same treatment group \citep{wang2014estimating}. 
This feature motivates the adoption of a mixed-effects ODE model to describe the time-course data, thereby enabling joint estimation of parameters at both the population (treatment group) and subject levels; e.g., if $Ka_j$ varies between subjects, it could be modeled as the sum of a fixed effect $Ka$ (for a given treatment group) and a random effect $b_{j, Ka}$ (for the between-subject variation in that treatment group), i.e., $Ka_j= Ka + b_{j, Ka}$,
where $b_{j, Ka}\sim N(0, \sigma_{Ka}^2)$. 

It is challenging to estimate parameters for mixed-effects ODE models in general. When there are closed-form solutions for the ODEs, the problem reduces to parameter inference for a standard nonlinear mixed-effects model
\citep{davidian2017nonlinear}. However, most non-linear ODEs lack closed-form solutions. Existing general methods for parameter inference in mixed-effects ODE models can be broadly categorized into those based on numerical integration, collocation, and Gaussian processes (GPs); these methods are reviewed in more detail in Section \ref{sec:review_me_ode}. Briefly, methods based on numerical solvers often suffer from convergence to local optima, while collocation methods tend to be sensitive to hyper-parameter choices. The idea of using manifold-constrained GPs has provided a promising approach for ODE inference \citep[MAGI,][]{yang2021inference}, but to our knowledge, the use of GPs to facilitate parameter inference for mixed-effects ODE
models remains to be explored.

Therefore, this paper develops an extension of manifold-constrained GPs to provide fast and accurate inference for mixed-effects ODEs.
The main challenge is to incorporate the hierarchical structure with subject-specific ODE parameters into the modeling framework in a computationally efficient way. To address this, we employ a normal approximation of the posterior distribution and maximize the marginal posterior, where the trajectory and random effects are integrated out. Our proposed method, manifold-constrained Gaussian process Inference for mixed-effects ODEs (MAGI-ME), performs favorably compared to other representative methods on simulated examples. We present an illustrative application on pharmacokinetic data, where two key practical advantages of our method are: (i) providing uncertainty quantification for the estimated subject-specific pharmacokinetic measures, such as peak plasma concentration and trough plasma concentration; (ii) readily generating predictions of plasma concentration at future times, which can be used to better assess therapeutic effectiveness and/or toxicity during the drug elimination phase. 

For the remainder of this paper, we begin with the background on mixed-effects ODE models and a review of relevant methods for parameter inference in Section \ref{sec:background}; since our work builds upon the MAGI GP-based framework, a review of the MAGI method for ODE inference is also provided. Section \ref{sec:method_chap_3} then presents our methodology: a Bayesian framework and associated optimization procedures for mixed-effects ODE inference. Section \ref{sec:simulation_chap_3} presents simulation studies that  assess the performance of MAGI-ME on a benchmark system and a more complex FitzHugh–Nagumo model \citep{fitzhugh1961impulses}. An application that fits and predicts the time-course plasma concentration of two treatment groups in the context of HIV therapy is presented in Section \ref{sec:application_chap_3}. We conclude the paper with a discussion in Section \ref{sec:discussion_chap_3}.

\section{Background}\label{sec:background}

\subsection{Mixed-effects ODE Models}
\label{sec:me_ode_form}
We consider mixed-effects ODE models described by the set of ODEs \citep[e.g.,][]{guedj2007maximum, prague2013nimrod,wang2014estimating,liu2019bayesian,clairon2024parameter}:
\begin{equation}
x_{ij}'(t) = f_i(\boldsymbol{x}_j(t),\boldsymbol{\theta}_{j},t),\quad t\in[0, T],i\in\{1,\cdots, m\}, j\in\{1,\cdots, s\},
    \label{eqn:mixed_effect_ode}
\end{equation}
where $m$ is the number of system components and $s$ is the number of subjects. We let $\boldsymbol{x}_j(t) = (x_{1j}(t),\cdots, x_{mj}(t))$ denote the trajectory of the $j-$th subject at time $t$, with each $x_{ij}(t)$ denoting the output of the $i-$th component for the $j-$th subject at time $t$. The functions $f_i(\cdot): \mathbb{R}^{m}\times \mathbb{R}^{l}\times \mathbb{R}\to \mathbb{R}$ are treated as known from the scientific context. Here, $\boldsymbol{\theta}_j=(\theta_{j1},\cdots, \theta_{jl})$ denotes a $l-$dimensional vector of unknown model parameters of the $j-$th subject. We assume $\boldsymbol{\theta}_j=\boldsymbol{\eta}+\boldsymbol{b}_j$, where $\boldsymbol{\eta}$ represents the vector of fixed effects and $\boldsymbol{b}_j$ represents the vector of random effects. A common assumption on the random effects is that $\boldsymbol{b}_j\sim N(\boldsymbol{0}, \boldsymbol{\Sigma}_{\boldsymbol{b}})$, where $\boldsymbol{\Sigma}_{\boldsymbol{b}}$ is the $l\times l$ covariance matrix of $\boldsymbol{b}_j$ \citep[]{wang2014estimating,liu2019bayesian,clairon2024parameter}. If we only have observations for one unit or subject, \eqref{eqn:mixed_effect_ode} reduces to an ODE model: $\boldsymbol{x}'(t) = \mathbf{f}(\boldsymbol{x}(t),\boldsymbol{\theta},t)$, where $\boldsymbol{x}(t)=(x_1(t),\cdots, x_m(t))$ and $\boldsymbol{\theta}$ is a $l-$dimensional vector of ODE model parameters.

Suppose measurements for each subject are taken at the discrete time points $\boldsymbol{\gamma}_j = (\boldsymbol{\gamma}_{1j},\cdots,\boldsymbol{\gamma}_{mj})$ with $\boldsymbol{\gamma}_{ij}= (\gamma_{ij1},\cdots, \gamma_{ijN_{ij}})$, i.e., the $i-$th component of the $j-$th subject is observed at $N_{ij}$ time points. Denoting the observations from the $j-$th subject by $\boldsymbol{y}_j(\boldsymbol{\gamma}_j)=(\boldsymbol{y}_{1j}(\boldsymbol{\gamma}_{1j}),\cdots, \boldsymbol{y}_{mj}(\boldsymbol{\gamma}_{mj}))$, we assume the subject-specific observations are subject to i.i.d.~additive Gaussian noise, i.e.,  \begin{equation}
\boldsymbol{y}_{ij}(\boldsymbol{\gamma}_{ij}) = \boldsymbol{x}_{ij}(\boldsymbol{\gamma}_{ij})+\boldsymbol{\epsilon}_{ij}(\boldsymbol{\gamma}_{ij}),\quad \boldsymbol{\epsilon}_{ij}(\boldsymbol{\gamma}_{ij}) \sim N(\mathbf{0}, \sigma_{ij}^2 \mathbf{I}_{N_{ij}}),
\label{eqn:likelihood}
\end{equation}
where $\sigma_{ij}^2$ is the variance of the $i$-th component of the $j$-th subject. Our main focus is inference of the fixed-effects parameter vector $\boldsymbol{\eta}$, covariance matrix of the random-effects $\boldsymbol{\Sigma}_{\boldsymbol{b}}$, and  individual trajectories $\boldsymbol{x}_j(t)$, given the noisy observations $\boldsymbol{y}_j(\boldsymbol{\gamma}_j)$ of each subject.

\subsection{Review of methods for mixed-effects ODE inference} 
\label{sec:review_me_ode}
The first broad category of inference methods uses numerical solvers, such as Euler and Runge–Kutta integrators \citep{yang2021inference}. 
\citet{bihorel2011scarabee} maximized the likelihood function of the observed data given the numerical solution to the ODEs. \citet{boersch2017debinfer} introduced a Bayesian framework by incorporating priors on the parameters and employing Markov chain Monte Carlo (MCMC) sampling for parameter inference. These numerical solver-based methods are generally applicable to mixed-effects ODE models, by constructing a likelihood function (or corresponding posterior distribution) that incorporates between-subject variation in the ODE parameters. 
However, the subject-level random effects increase the dimension of the parameter space and the associated computational burden of numerical solvers increases linearly with the number of subjects.
For better scalability, \citet{guedj2007maximum} developed a maximum likelihood approach that employs a Newton-like algorithm with Gaussian quadrature to integrate out the random effects. In an analogous Bayesian approach, \citet{prague2013nimrod} used a normal approximation to the posterior distribution for deriving \textit{maximum a posteriori} (MAP) estimates that bypass the need for MCMC sampling. However, convergence difficulties are common to numerical solver-based methods of inference, as the parameter space can have many local optima when the ODE solutions are highly sensitive to the parameters \citep{liang2008parameter}. 

Collocation methods approximate ODE solutions via spline basis expansions, thereby bypassing numerical solvers. 
\citet{varah1982spline} first proposed a two-step collocation method that smooths noisy data with splines and then minimizes the discrepancy between spline derivatives and ODEs. To avoid an inaccurate estimation of $\boldsymbol{x}(t)$ in the first step, 
\citet{ramsay2007param} pioneered a generalized profiling procedure that optimizes the spline coefficients and ODE parameters together using a penalized likelihood. \citet{wang2014estimating} adapted this method to the context of the mixed-effects ODE model, using a three-level nested optimization procedure.
\citet{liu2019bayesian} proposed a Bayesian approach using MCMC to estimate the mixed-effects parameters, introducing a hierarchical structure that allows for different distributions of measurement errors.
However, these methods tend to lack general guidelines for hyper-parameter selection (which influences the accuracy of collocation-based inference) and for uncertainty quantification on a subject-level. 

Gaussian processes (GPs) provide an alternative way to bypass numerical solvers within a Bayesian paradigm. In the context of ODEs (without random effects), these methods impose a GP on prior on $\boldsymbol{x}(t)$, allowing for a closed-form expression of the conditional distribution of $\boldsymbol{x}'(t)|\boldsymbol{x}(t)$. The main challenge is to address the conceptual incompatibility of $\boldsymbol{x}'(t)$ between the GP (via $p(\boldsymbol{x}'(t))|\boldsymbol{x}(t)$) and ODE (via the function $\mathbf{f}$) specifications. Early work \citep{calderhead2008accelerating,dondelinger2013ode} proposed gradient matching along with heuristics to combine the two different specifications of $\boldsymbol{x}'(t)$, which was subsequently refined by \citet{wenk2019fast}. \citet{yang2021inference} proposed the manifold-constrained Gaussian process Inference (MAGI) method for ODEs, which addressed the incompatibility by conditioning the GP for  $\boldsymbol{x}(t)$ on the manifold where the ODEs specified by $\mathbf{f}$ must be satisfied. MAGI provides fast and accurate parameter estimation for ODEs from noisy and sparse observations \citep{wong2023estimating}. 
Further extensions of MAGI have incorporated the historical outputs in delay differential equations \citep{zhao2024inference} and time-varying parameters \citep{sun2023manifold}, which demonstrate  the efficacy of using manifold constraints to facilitate fitting more complex dynamic systems. 
However, it is not straightforward to apply the MAGI framework for fitting a mixed-effects ODE model. 
First, MAGI assumes independent GP priors for each component of the differential equation and is designed for estimating ODE parameters from a single observation trajectory. Mixed-effects modeling requires handling multiple trajectories and potential correlation among subjects. 
Second, the parameter space grows linearly with the number of subjects, making MCMC methods burdensome for posterior sampling. A common technique in nonlinear mixed-effects models is to integrate over the random effects \citep{pinheiro1995approximations,ke2001semiparametric}. Here, an efficient procedure for performing marginalization over the random effects within the MAGI framework is needed. 

\subsection{Review of the MAGI method for ODE inference}
\label{sec:review_magi}
We briefly review the MAGI method \citep{yang2021inference} for parameter inference in the ODE model $\boldsymbol{x}'(t) = \mathbf{f}(\boldsymbol{x}(t),\boldsymbol{\theta},t)$. 
MAGI imposes an independent GP prior on each component $x_i(t)$ such that $x_i(t)\sim \mathcal{GP}(\mu_i, \mathcal{K}_i), t\in[0, T],$
 where $\mathcal{K}_i:\mathbb{R}\times \mathbb{R}\to \mathbb{R}$ is a positive definite covariance kernel and $\mu_i:\mathbb{R} \to \mathbb{R}$ is the mean function, typically set as $\mu_i(t) \equiv 0$. A general prior $\pi(\cdot)$ is placed on the model parameters $\boldsymbol{\theta}$. In this setting, the noisy observations may be denoted as $\boldsymbol{y}(\boldsymbol{\gamma}) = \left(\boldsymbol{y}_1(\boldsymbol{\gamma}_1),\cdots, \boldsymbol{y}_m(\boldsymbol{\gamma}_m)\right)$, where $\boldsymbol{\gamma} = (\boldsymbol{\gamma}_1,\cdots, \boldsymbol{\gamma}_m)$ is the collection of observation time points, and MAGI assumes that  $\boldsymbol{y}_i(\boldsymbol{\gamma}_i) = \boldsymbol{x}_i(\boldsymbol{\gamma}_i) + \boldsymbol{\epsilon}_i(\boldsymbol{\gamma}_i)$, where $\boldsymbol{\epsilon}_i(\boldsymbol{\gamma}_i)\sim N(\boldsymbol{0}, \sigma_i\cdot \mathbf{I}_{|\boldsymbol{\gamma}_i|})$.

The GP prior on $\boldsymbol{x}(t)$ provides a fully-specified GP form for the conditional distribution of $\boldsymbol{x}'(t)$ given $\boldsymbol{x}(t)$. To resolve the conceptual incompatibility between the GP and ODE specifications of $\boldsymbol{x}'(t)$, the MAGI method links this GP-specified distribution of $\boldsymbol{x}'(t)$ with the function $\mathbf{f}$ in the ODE by introducing a random variable $W$ which measures the uniform deviation between them, i.e.,  $W = \sup_{t\in[0,T], i\in\{1,\cdots, m\}} \big|x'_i(t)-f_i(\boldsymbol{x}(t),  \boldsymbol{\theta},t)\big|$. Conditioning on the manifold constraint $W=0$ thus ensures that the ODE must be satisfied. In practice, $W$ cannot be computed directly; instead, it is approximated by taking the uniform deviation over a finite set of discretization points $\boldsymbol{I} = \{t_1,\cdots, t_n\}$ such that $\boldsymbol{\gamma}\subset \boldsymbol{I}\subset [0,T]$ and defining  $W_{\boldsymbol{I}} = \sup_{t\in\boldsymbol{I}, i\in\{1,\cdots, m\}} \big|x'_i(t)-f_i\left(\boldsymbol{x}(t), \boldsymbol{\theta},t\right)\big|.$ Therefore, the computable joint posterior of $\boldsymbol{\theta}$ and $\boldsymbol{x}(\boldsymbol{I})$ conditional on $\boldsymbol{W}_{\boldsymbol{I}} = \boldsymbol{0}$ and the noisy measurements $\boldsymbol{y}(\boldsymbol{\gamma})$ is given by 
\begin{equation}
    \begin{aligned}
    &p\left(\boldsymbol{\theta}, \boldsymbol{x}(\boldsymbol{I})| \boldsymbol{W}_{\boldsymbol{I}} = \boldsymbol{0}, \boldsymbol{y}(\boldsymbol{\gamma})\right)
    \propto p\left(\boldsymbol{\theta}, \boldsymbol{x}(\boldsymbol{I}), \boldsymbol{W}_{\boldsymbol{I}} = \boldsymbol{0},\boldsymbol{y}(\boldsymbol{\gamma})\right)\\
    &=\pi(\boldsymbol{\theta}) \times p\left(\boldsymbol{x}(\boldsymbol{I})\right) \times p\left(\boldsymbol{y}(\boldsymbol{\gamma})| \boldsymbol{x}(\boldsymbol{I})\right)\times p\left(\boldsymbol{x}'(\boldsymbol{I})=\mathbf{f}(\boldsymbol{x}(\boldsymbol{I}), \boldsymbol{\theta}, \boldsymbol{I}) |\boldsymbol{x}(\boldsymbol{I})\right).
\end{aligned} 
\label{eqn:magi_posterior}
\end{equation}

Hamiltonian Monte Carlo \citep[HMC,][]{neal2011mcmc} is used to draw samples of $\boldsymbol{\theta}$ and $\boldsymbol{x}(\boldsymbol{I})$ from this posterior. Denser discretization sets $\boldsymbol{I}$ provide a more accurate approximation of the manifold constraint at the cost of computation time \citep{wong2022magi}, and also increase the contributions of the terms in \eqref{eqn:magi_posterior} involving $\boldsymbol{I}$, namely $p\left(\boldsymbol{x}(\boldsymbol{I})\right)$ and $p\left(\boldsymbol{x}'(\boldsymbol{I}) = \mathbf{f}(\boldsymbol{x}(\boldsymbol{I}), \boldsymbol{\theta}, \boldsymbol{I}) |\boldsymbol{x}(\boldsymbol{I})\right)$, while the likelihood term $p \left(\boldsymbol{y}(\boldsymbol{\gamma})| \boldsymbol{x}(\boldsymbol{I})\right)$ does not change. Hence, the MAGI method introduced a tempering hyper-parameter $\lambda = m|\boldsymbol{I}|/\sum_{i=1}^m |\boldsymbol{\gamma}_i|$ (i.e., the total number of discretization points divided by the total number of observation points) to maintain the balance between the GP prior and the likelihood across different cardinalities of $|\boldsymbol{I}|$. The terms associated with the
GP are tempered as $\left[p\left(\boldsymbol{x}(\boldsymbol{I})\right)p\left(\boldsymbol{x}'(\boldsymbol{I}) =\mathbf{f}(\boldsymbol{x}(\boldsymbol{I}), \boldsymbol{\theta}, \boldsymbol{I}) |\boldsymbol{x}(\boldsymbol{I})\right) \right]^{1/{\lambda}}$. It becomes apparent that scalability challenges will be more pronounced in mixed-effects ODEs, as the dimension of the posterior 
will increase linearly with both $|\boldsymbol{I}|$ and the number of subjects $s$.

\section{Methodology}
\label{sec:method_chap_3}

\subsection{Bayesian Framework}

We impose a GP prior on each component for the $j-$th subject $x_{ij}(t)$ such that 
\begin{equation}
    x_{ij}(t)\sim \mathcal{GP}(\mu_{ij}, \mathcal{K}_{ij}),\quad t\in[0,T].
\end{equation}
This facilitates a GP form for the conditional distribution of $\boldsymbol{x}_j'(t)$ given $\boldsymbol{x}_j(t)$. Since $\boldsymbol{b}_j$ and $\boldsymbol{\eta}$ jointly determine $\boldsymbol{\theta}_j$, we can rewrite  \eqref{eqn:mixed_effect_ode} as $x'_{ij}(t) = f_i(\boldsymbol{x}_j(t),\boldsymbol{\eta},\boldsymbol{b}_{j},t)$, where $\boldsymbol{\eta}$ is the vector of fixed effects, and $\boldsymbol{b}_j$ is the vector of random effects assumed to follow $N(\boldsymbol{0}, \boldsymbol{\Sigma}_{\boldsymbol{b}})$, i.e., $\boldsymbol{\Sigma}_{\boldsymbol{b}}$ is the covariance matrix of $\boldsymbol{b}_j$.
To link this GP of $\boldsymbol{x}_j'(t)|\boldsymbol{x}_j(t)$ with the mixed-effects ODE model structure, we define the subject-specific random variable $W_j$ which measures the uniform deviation between the ODE and the stochastic process as 
$W_j = \sup_{t\in[0,T], i\in\{1,\cdots, m\}} \big|x'_{ij}(t)-f_{i}(\boldsymbol{x}_j(t), \boldsymbol{\eta}, \boldsymbol{b}_j,t)\big|.$
We approximate $W_j$ by a finite set of $n_j$ discretization points $\boldsymbol{I}_j = \{t_1,t_2,\cdots, t_{n_j}\}$, where $\bigcup_{i=1}^m\boldsymbol{\gamma}_{ij} \subset \boldsymbol{I}_j$, i.e.,
$W_{\boldsymbol{I}_j} = \max_{t\in\boldsymbol{I}_j, i\in\{1,\cdots, m\}} \big|x'_{ij}(t)-f_{i}(\boldsymbol{x}_j(t), \boldsymbol{\eta},\boldsymbol{b}_j,t)\big|.$

Unlike the ODE setting in Section \ref{sec:review_magi}, where $\boldsymbol{I} $ denotes a single discretization set, here, $\boldsymbol{I}=(\boldsymbol{I}_1,\cdots, \boldsymbol{I}_s)$ is the collection of discretization sets corresponding to all $s$ subjects. 
Similarly, we define  $\boldsymbol{x}(\boldsymbol{I})=(\boldsymbol{x}_1(\boldsymbol{I}_1),\cdots, \boldsymbol{x}_s(\boldsymbol{I}_s))$, $\boldsymbol{b} = (\boldsymbol{b}_1,\cdots, \boldsymbol{b}_s)$,  $\boldsymbol{\gamma} = (\boldsymbol{\gamma}_1,\cdots, \boldsymbol{\gamma}_s)$, 
and $\boldsymbol{W}_{\boldsymbol{I}} = (W_{\boldsymbol{I}_1},\cdots,W_{\boldsymbol{I}_s})$. 
To complete the hierarchical structure
for modeling the variation of subject-specific ODE parameters, we place a general prior $\pi(\cdot)$ on $\boldsymbol{\eta}$ and $\boldsymbol{\Sigma}_{\boldsymbol{b}}$.
To satisfy all of the subject-specific ODEs, the manifold constraint is approximated by setting $\boldsymbol{W}_{\boldsymbol{I}} = \boldsymbol{0}$. 
The full posterior distribution is comprised of $\boldsymbol{\eta}$, $\boldsymbol{\Sigma}_{\boldsymbol{b}}$, $\boldsymbol{b}$ and $\boldsymbol{x}(\boldsymbol{I})$ given $\boldsymbol{W}_{\boldsymbol{I}} = \boldsymbol{0}$ and $\boldsymbol{y}(\boldsymbol{\gamma})$, which as detailed in  Section A.1 of the Appendix, can be written and factorized as
\begin{equation}
\begin{aligned}
&p\left(\boldsymbol{\eta},\boldsymbol{\Sigma}_{\boldsymbol{b}},\boldsymbol{b},  \boldsymbol{x}(\boldsymbol{I}), \boldsymbol{W}_{\boldsymbol{I}} = \boldsymbol{0}, \boldsymbol{y}(\boldsymbol{\gamma})\right)=\underbrace{\pi(\boldsymbol{\eta})}_{(1)}\times \underbrace{\pi(\boldsymbol{\Sigma}_{\boldsymbol{b}})}_{(2)}\times\underbrace{p(\boldsymbol{b}|\boldsymbol{\Sigma}_{\boldsymbol{b}})}_{(3)}\times \underbrace{p(\boldsymbol{x}(\boldsymbol{I}))}_{(4)}
\times \underbrace{p(\boldsymbol{y}(\boldsymbol{\gamma})|\boldsymbol{x}(\boldsymbol{I}))}_{(5)}\\
&\times \underbrace{p(\boldsymbol{x}'(\boldsymbol{I}) = \mathbf{f}(\boldsymbol{x}(\boldsymbol{I}), \boldsymbol{\eta}, \boldsymbol{b},\boldsymbol{I})|\boldsymbol{x}(\boldsymbol{I}))}_{(6)}.
\end{aligned}
\label{eqn:joint_post_bayes_chap_3}
\end{equation}

The first and second term are the prior densities of $\boldsymbol{\eta}$ and $\boldsymbol{\Sigma}_{\boldsymbol{b}}$ respectively. The third term represents the between-subject variation in the ODE parameters, usually assumed to follow a multivariate normal distribution. The fourth term is the multivariate normal density of $\boldsymbol{x}(\boldsymbol{I})$ based on the GP prior.  The fifth term is the normal likelihood of the noisy observations. Note that a different distribution for the measurement errors can be easily incorporated in our method by changing this likelihood function. The sixth term is the multivariate normal density for the conditional distribution of $\boldsymbol{x}'(\boldsymbol{I})|\boldsymbol{x}(\boldsymbol{I})$ evaluated at $\boldsymbol{x}'(\boldsymbol{I}) = \mathbf{f}(\boldsymbol{x}(\boldsymbol{I}),\boldsymbol{\eta},\boldsymbol{b},\boldsymbol{I} )$ provided that the covariance kernel $\mathcal{K}$ is associated with twice-differentiable curves, i.e., 
 $\boldsymbol{x}'_{ij}(\boldsymbol{I}_j)|\boldsymbol{x}_{ij}(\boldsymbol{I}_j)\sim N\left(\boldsymbol{\mu}'_{ij}(\boldsymbol{I}_j)+\boldsymbol{m}_{ij}(\boldsymbol{x}_{ij}(\boldsymbol{I}_j)-\boldsymbol{\mu}_{ij}(\boldsymbol{I}_j)),\boldsymbol{\zeta}_{ij}\right)$,
where   $\boldsymbol{m}_{ij}   ={ }^{\prime} \mathcal{K}_{ij}(\boldsymbol{I}_j, \boldsymbol{I}_j) \mathcal{K}_{ij}(\boldsymbol{I}_j, \boldsymbol{I}_j)^{-1} $ and $\boldsymbol{\zeta}_{ij} =\mathcal{K}_{ij}^{\prime \prime}(\boldsymbol{I}_j, \boldsymbol{I}_j)-{ }^{\prime} \mathcal{K}_{ij}(\boldsymbol{I}_j, \boldsymbol{I}_j) \mathcal{K}_{ij}(\boldsymbol{I}_j, \boldsymbol{I}_j)^{-1} \mathcal{K}^{\prime}_{ij}(\boldsymbol{I}_j, \boldsymbol{I}_j)$ with
 ${ }^{\prime} \mathcal{K}_{ij}=\frac{\partial}{\partial s} \mathcal{K}_{ij}(s, t), \mathcal{K}^{\prime}_{ij}=\frac{\partial}{\partial t} \mathcal{K}_{ij}(s, t) \text {, and } \mathcal{K}^{\prime \prime}_{ij}=\frac{\partial^2}{\partial s \partial t} \mathcal{K}_{ij}(s, t)$. The closed-form expressions for each term of \eqref{eqn:joint_post_bayes_chap_3} are provided in Section A.1 of the Appendix.

In practice, we choose the Matern class where the covariance of the $i$-th component of the $j-$th subject between time points $s$ and $t$ is given by $\mathcal{K}_{ij}(s, t)=\phi_{ij,1} \frac{2^{1-\nu}}{\Gamma(\nu)}\left(\sqrt{2 \nu} \frac{d}{\phi_{ij,2}}\right)^\nu B_\nu\left(\sqrt{2 \nu} \frac{d}{\phi_{ij,2}}\right),$ where $ d=|s-t|$, $\Gamma$ is the Gamma function, $B_\nu$ is the modified Bessel function of the second kind, and $\nu$ is the degree of freedom. $\mathcal{K}$ is $k-$times differentiable if and only if $\nu>k$ \citep{williams2006gaussian}. Typical choices for $\nu$ are $2.01$ or $2.5$, which ensure the twice-differentiability; $\nu=2.01$ is the default choice that is suitable for rougher curves, while $\nu=2.5$ is adequate for smoother curves and has faster computation speed \citep{wong2022magi}. $\mathcal{K}_{ij}(s,t)$ has two hyper-parameters, $\phi_{ij,1}$ and $\phi_{ij,2}$, that respectively control the overall variance and bandwidth of the $i$-th component of the $j-$th subject. 

\subsection{Posterior Inference}
\label{sec:posterior_inference}
After completing the choice of priors, Bayesian inference could be performed by drawing MCMC samples from the posterior distribution in \eqref{eqn:joint_post_bayes_chap_3}, as in the original MAGI method \citep{yang2021inference}. However, in the mixed-effects setup, even a moderate number of subjects will entail a high-dimensional  $\boldsymbol{x}(\boldsymbol{I})$ to adequately represent their trajectories, e.g., $|\boldsymbol{x}(\boldsymbol{I})| >2000$ as in the application (Section \ref{sec:application_chap_3}). 
Instead, for computational efficiency we shall perform optimization using a normal approximation to the posterior distribution.
We denote $\boldsymbol{\Sigma}_{\boldsymbol{b}}$ as  $\boldsymbol{\Sigma}_{\boldsymbol{b}}(\boldsymbol{\beta})$, where $\boldsymbol{\beta}$ is the vectorized representation of $\boldsymbol{\Sigma}_{\boldsymbol{b}}$. Since the covariance matrix $\boldsymbol{\Sigma}_{\boldsymbol{b}}(\boldsymbol{\beta})$ is positive-definite, we use the Cholesky factorization to represent  
$\boldsymbol{\Sigma}_{\boldsymbol{b}}(\boldsymbol{\beta}) = \boldsymbol{B}(\boldsymbol{\beta}) [\boldsymbol{B}(\boldsymbol{\beta})]^\top$ where $\boldsymbol{B}(\boldsymbol{\beta})$ is lower-triangular, which
transforms a constrained optimization problem to an unconstrained one. 
Define 
$\boldsymbol{\omega} = (\boldsymbol{\beta}, \boldsymbol{\eta})^\top$ (i.e., a low-dimensional parameter vector for $\boldsymbol{\Sigma}_{\boldsymbol{b}}$ and the fixed-effects parameters) and $\boldsymbol{u} = (\boldsymbol{b},\boldsymbol{x}(\boldsymbol{I}))^\top$ 
(i.e., a high-dimensional vector with the random effects and trajectories for all subjects). 
As the prior distribution for $\boldsymbol{x}(\boldsymbol{I})$ is multivariate normal, its posterior also tends to be close to Gaussian \citep{gelman1995bayesian}; meanwhile, empirical evidence suggests the posteriors for the ODE parameters in MAGI tend to be reasonably normal  \citep{wong2023estimating}. This motivates the normal approximation
\begin{equation}p(\boldsymbol{u},\boldsymbol{\omega}| \boldsymbol{W}_{\boldsymbol{I}} = \boldsymbol{0}, \boldsymbol{y}(\boldsymbol{\gamma})) \approx N(\hat{\boldsymbol{\xi}}, [I(\hat{\boldsymbol{\xi}})]^{-1}),
    \label{eqn:normality_assumption}
\end{equation}
where $\boldsymbol{\xi} = (\boldsymbol{u},\boldsymbol{\omega})$, $\hat{\boldsymbol{\xi}} = \arg\max_{\boldsymbol{\xi}}    p(\boldsymbol{\xi}|\boldsymbol{W}_{\boldsymbol{I}} = \boldsymbol{0}, \boldsymbol{y}(\boldsymbol{\gamma}))$ and $I(\boldsymbol{\xi}) = -\frac{d^2}{d\boldsymbol{\xi}\boldsymbol{\xi}^\top}\log p(\boldsymbol{\xi}| \boldsymbol{W}_{\boldsymbol{I}} = \boldsymbol{0}, \boldsymbol{y}(\boldsymbol{\gamma}))$. 
However, maximizing \eqref{eqn:normality_assumption} directly can still be challenging due to the dimensionality of  $\boldsymbol{\xi}$.

Thus, we propose using two nested levels of optimization for estimating $\boldsymbol{\omega}$ and $\boldsymbol{u}$. 
In the inner level, we maximize the posterior distribution with respect to $\boldsymbol{u}$ given $\boldsymbol{\omega}$. In the outer level, we integrate out $\boldsymbol{u}$ from the posterior distribution and maximize the Laplace approximation of the marginal posterior with respect to $\boldsymbol{\omega}$. Details of the nested optimization procedure and the derivation to obtain the standard error of $\boldsymbol{u}$ are provided in Section A.2 and Section A.3 of the Appendix, respectively.

\subsection{Practical Implementation}
\label{sec:Practical Implementation}
The practical steps to implement the MAGI-ME method are provided as follows. 
First, we fit a GP to the noisy observations $\boldsymbol{y}_{ij}(\boldsymbol{\gamma}_{ij})$ for each component $i$ of the $j-$th subject. If $\sigma_{ij}$ is known, we maximize the marginal likelihood $p(\phi_{ij,1},\phi_{ij,2}|\boldsymbol{y}_{ij}(\boldsymbol{\gamma}_{ij}))$; if $\sigma_{ij}$ is unknown, $(\phi_{ij,1},\phi_{ij,2})$ can be obtained by maximizing the marginal likelihood $p(\phi_{ij,1},\phi_{ij,2}, \sigma^2_{ij}|\boldsymbol{y}_{ij}(\boldsymbol{\gamma}_{ij}))$. The hyper-parameter values $\phi_{ij,1},\phi_{ij,2}$ are held fixed during the subsequent optimizations. 
Second, we set the starting parameter values for the nested optimization procedure.
For $\boldsymbol{x}(\boldsymbol{I})$, observations $\boldsymbol{y}(\boldsymbol{\gamma})$ are used as starting values for $\boldsymbol{x}(\boldsymbol{\gamma})$, and the mean of the GP fit to the noisy observations is used for time points $\boldsymbol{I}\backslash \boldsymbol{\gamma}$. Starting values for subject-specific ODE parameters $\boldsymbol{\theta}_j$ are obtained by optimizing the posterior in \eqref{eqn:normality_assumption}; based on these, sensible starting values for the mixed-effects parameters can be computed: ${\boldsymbol{\eta}}=\sum_{j=1}^s \boldsymbol{\theta}_j$, ${\boldsymbol{b}}_j =\boldsymbol{\theta}_j-\boldsymbol{\eta}$, and $\boldsymbol{\beta}$ is obtained from the sample covariance matrix of $\boldsymbol{b}$.

Then, we define the posterior distribution \eqref{eqn:normality_assumption} as a \verb!C++! template and pass it to `TMB' to obtain the MAP estimates and standard errors for all the elements of $\boldsymbol{\omega}$ and $\boldsymbol{u}$. 
The MAP estimates of $\boldsymbol{\eta}, \boldsymbol{\Sigma}_{\boldsymbol{b}}$, and $\boldsymbol{x}(\boldsymbol{I})$ are taken to be the estimated mixed-effects parameters and inferred trajectories.
TMB is an R package developed for fast implementation of complex nonlinear random effects models \citep{JSSv070i05}. TMB performs a two-level nested optimization, which we adapt to carry out the inner and outer optimizations proposed in Section \ref{sec:posterior_inference}. 
The outer-level optimization to obtain $\boldsymbol{\omega}$ involves the log-determinant of the covariance matrix of $\boldsymbol{u}$,
for which we leverage the Cholesky decomposition code by \citet{chen2008algorithm}. TMB's built-in automatic differentiation enables the Laplace approximation in the outer optimization step and the gradient for $\boldsymbol{\omega}$ to be efficiently obtained. 

An important practical consideration  is selecting the discretization set $\boldsymbol{I}_j$ for the $j-$th subject. As suggested by \citet{zhao2024inference}, $\boldsymbol{I}_j$ needs to be sufficiently dense to infer a smooth system trajectory. However, this rule-of-thumb can be more difficult to apply when there is a large number of subjects and trajectories to examine. As a more systematic approach, we can follow the general guidelines in \citet{wong2022magi}. We begin by taking $\boldsymbol{I}_{0,j}$ as the smallest evenly-spaced set that includes all the observation time points $\boldsymbol{\gamma}_j$. Then, we construct subsequent sets $\boldsymbol{I}_{j,k}\supset \boldsymbol{I}_{j,k-1}, k\ge 1$ by inserting one equally-spaced discretization point between each adjacent pair of points in $\boldsymbol{I}_{j,k-1}$, and stop when the estimates have stabilized. Furthermore, MAGI-ME can fit the data and generate future predictions simultaneously, by constructing $\boldsymbol{I}_j$ to include future time points $\{t\in\boldsymbol{I}_j| t>\max (\boldsymbol{\gamma}_j)\}$. The inferred trajectory and corresponding point-wise credible interval for $\boldsymbol{x}(\boldsymbol{I}_j)$ then includes the predictions for the future time points of interest. Credible intervals for pharmacokinetic measures can likewise be easily obtained using the standard error of $\boldsymbol{u}$, which contains point-wise standard errors for the inferred trajectory.    

We numerically validate our proposed method and implementation by comparing it with the posterior means obtained from MCMC samples drawn from the posterior distribution in \eqref{eqn:joint_post_bayes_chap_3}, under varying discretization levels $|\boldsymbol{I}_j|=21,41,81$ with $s=20$ subjects. For this purpose, we used a population growth model with a known analytic solution (see model description in Section A.4.1 of the Appendix). The results, shown in Section A.4.2 of the Appendix and Table S1, indicate that MAGI-ME yields comparable parameter estimates and quality of credible intervals, while being 2–3 orders of magnitude faster, when compared to MCMC sampling from the full posterior distribution. 
To provide further validation, we also compared our results to those obtained using the \texttt{nlme()} function in R, a widely used tool to fit linear and non-linear mixed-effects models \citep{Pinheiro2021-ui}. Note that the \texttt{nlme()} function is limited to ODEs with analytic solutions, as it requires the model for the underlying trajectory to be explicitly specified as analytic ODE solutions. MAGI-ME produces similar results to the \texttt{nlme()} function (see Section A.4.3 of the Appendix and Table S2), which provides further support for the quality of our inference.

\section{Simulation Study}
\label{sec:simulation_chap_3}

\subsection{Benchmark Model}
\label{sec:simulation_benchmark}

In previous studies, the performance of different methods for mixed-effects ODE inference was typically assessed using simple homogeneous ODE models with solution trajectories that stabilized at a steady state \citep{wang2014estimating,liu2019bayesian,clairon2024parameter}. As a more challenging example, we consider a nonhomogeneous ODE exhibiting oscillatory behavior in its solution trajectory, which is a modified form of the van der Pol equation:\begin{equation}
    x_{j}'(t) = \theta_{j1}(1-x^2_j(t))\cdot x_j(t)-\theta_{j2}\sin(t),
\label{eqn:ode_simulation}
\end{equation} where $\theta_{j1}$ determines the intensity of the nonlinear feedback in the system, and $\theta_{j2}$ determines the amplitude of the periodic forcing term $\sin(t)$. This setup leads to oscillations with varying amplitudes, which can be difficult for methods based on splines and GPs to approximate well. 
Moreover, this ODE does not have a closed-form analytic solution, and so the $\mathtt{nlme()}$ function is inapplicable for performing inference.

We set the true fixed effects as $\eta_1=\eta_2 = 0.6$ and generate the subject-specific ODE parameters $\boldsymbol{\theta}_j = 
\begin{bsmallmatrix}
\theta_{j1}\\
\theta_{j2}
\end{bsmallmatrix} = \begin{bsmallmatrix}
\eta_1\\
\eta_2
\end{bsmallmatrix} + \begin{bsmallmatrix}
b_{j1}\\
b_{j2}
\end{bsmallmatrix}$ with $\begin{bsmallmatrix}b_{j1}\\
b_{j2}
\end{bsmallmatrix}\sim N\left(\begin{bsmallmatrix}
0\\
0
\end{bsmallmatrix},\boldsymbol{\Sigma}_{\boldsymbol{b}} \right)$ where 
$ \boldsymbol{\Sigma}_{\boldsymbol{b}} =  
\begin{bsmallmatrix}
   0.01 & 0.01\\
    0.01 & 0.01
\end{bsmallmatrix}$ for $N = 25$ subjects. The initial condition $x_j(0)$ is independently generated for each subject from the normal distribution $N(1, \sigma_0^2)$. To create the simulation data, we numerically solve the ODE defined in \eqref{eqn:ode_simulation} for each subject over 21 equally-spaced observation time points in $[0,20]$ using the R package `deSolve' \citep{soetaert2010solving}. The observations  are then generated by adding i.i.d.~$N(0, \sigma^2)$ noise to the ODE solution at each time point. We set $\sigma_0 = \sigma = 0.03$ and repeat the above data generating procedure to create 100 simulated datasets. 

We compare MAGI-ME with other representative methods for mixed-effects ODE inference based on numerical solvers and collocation in a Bayesian setting. 
We consider two approaches for numerical solver-based inference using the \texttt{rstan} R package \citep{carpenter2017stan}: (1) maximizing the joint posterior distribution, referred to as `Numerical-MAP', and (2) sampling from the joint posterior via MCMC, referred to as `Numerical-MCMC'. 
To mitigate convergence difficulties for `Numerical-MAP', we run 20 tries from different random starting parameter values and select the best parameter set based on log-posterior values. 
For the collocation method, 
we employ the implementation of \citet{liu2019bayesian}, which draws MCMC samples of the parameters and trajectories. For fair comparison, we place the same diffuse priors on the model parameters for all the methods; priors and full implementation details of each method are provided in Section B.1 of the Appendix.

We use two performance metrics to assess the quality of parameter estimates and trajectory recovery. First, we evaluate the accuracy of parameter estimates by computing the root mean squared error (RMSE) of the estimated parameters to the true parameter values. Second, we use the trajectory mean squared error (MSE), as in \citet{liu2019bayesian}, to assess how well the ODE solution trajectory is recovered from the parameter estimates. The trajectory MSE is calculated by the following steps: (i) we use the numerical solver to compute the true trajectory  $\boldsymbol{C}(\boldsymbol{\gamma}_j)$ based on the true $\boldsymbol{\theta}_j$ and $x_j(0)$ for each subject over the observation time points $\boldsymbol{\gamma}_j$; (ii) we use the numerical solver to reconstruct the trajectory $\hat{\boldsymbol{C}}(\boldsymbol{\gamma}_j)$ implied by the method's estimated parameter values; (iii) we compute the MSE of the reconstructed trajectory to the true trajectory and average over all subjects:
$MSE = \frac{1}{s}\sum_{j=1}^s \frac{1}{|\boldsymbol{\gamma}_j|}\left\|\boldsymbol{C}(\boldsymbol{\gamma}_j)-\hat{\boldsymbol{C}}(\boldsymbol{\gamma}_j)\right\|^2.$

We compute the parameter RMSEs and trajectory MSEs across the 100 simulated datasets, as summarized in Table \ref{tab:param_simulation_1}. 
Among these methods, MAGI-ME showcases favorable performance in recovering the parameters and system trajectories.
As illustrated in Figure S1 from Section B.2 of the Appendix, MAGI-ME well-recovers the true trajectory and has a narrow 95$\%$ credible interval around the truth. 
Numerical-MCMC often struggles to produce reasonable parameter estimates and tends to yield high average trajectory MSEs. 
In contrast, the presence of local optima is largely mitigated by Numerical-MAP with different random starting points, but it still does not reliably provide parameter inference as incorrect estimates persist in a few simulated data sets.  
The collocation method exhibits some bias in recovering the system, notably in the variance components of $\boldsymbol{\Sigma}_{\boldsymbol{b}}$ and reconstructed trajectories. Moreover, MAGI-ME is the fastest among all the methods. 
\begin{table}[hbt!]
\caption{Average fixed-effects parameter estimates (with parameter RMSEs) for the mixed-effects model in \eqref{eqn:ode_simulation}, along with average trajectory MSE across 100 simulated data sets using the different methods. $\boldsymbol{\Sigma}_{\boldsymbol{b}}(a,b)$ denotes the $(a,b)$-th entry of $\boldsymbol{\Sigma}_{\boldsymbol{b}}$. The last column gives the average runtime (in minutes, on a single CPU core).}
\resizebox{\textwidth}{!}{
\begin{tabular}{@{}crrrrrrrrrrrrr@{}}
\toprule
\multirow{2}{*}{Method} 
& \multicolumn{2}{c}{$\eta_1$} 
& \multicolumn{2}{c}{$\eta_2$} 
& \multicolumn{2}{c}{$\boldsymbol{\Sigma}_{\boldsymbol{b}}(1,1)$}
& \multicolumn{2}{c}{$\boldsymbol{\Sigma}_{\boldsymbol{b}}(1,2)$}
& \multicolumn{2}{c}{$\boldsymbol{\Sigma}_{\boldsymbol{b}}(2,2)$}
& \multirow{2}{*}{MSE}
& \multirow{2}{*}{Runtime}\\ 
\cmidrule(lr){2-3} \cmidrule(lr){4-5} \cmidrule(lr){6-7} \cmidrule(lr){8-9} \cmidrule(lr){10-11}
& Est & RMSE & Est & RMSE & Est & RMSE & Est & RMSE & Est & RMSE &  \\ 
\midrule
MAGI-ME &  0.5790  & 0.0284 & 0.5897 & 0.0216 & 0.0091 & 0.0026 &0.0092 &0.0025  & 0.0094& 0.0025 & 0.0002 & 4.37 \\
Collocation &   0.5168        & 0.0842  &0.5700&  0.0336&0.0033 &0.0069& 0.0008&0.0093&0.0057&0.0047&0.0016& 26.31\\
Numerical-MAP   &  0.6351   & 0.3957 &   0.6115  &   0.1464         & 0.0077&  0.0084&0.0066 & 0.0041&0.0080  & 0.0072 & 0.0004&8.90 \\
Numerical-MCMC &7.0298&8.0727&2.7333&2.7405&0.3928&1.2922&-0.0143&0.1317&0.3142&1.0707& 0.1065& 1652.36 \\
\bottomrule
\end{tabular}
}
\label{tab:param_simulation_1}
\end{table}

\subsection{FitzHugh–Nagumo Equations}
\label{sec:simulation_fn}
Previous methods for mixed-effects ODE inference were typically assessed under simple mixed-effects ODE models involving only one component per subject \citep{wang2014estimating,liu2019bayesian,liu2021semiparametric}. In this section, we demonstrate MAGI-ME's capability to recover a multi-component mixed-effects ODE model using the FitzHugh–Nagumo (FN) equations. Originally written to describe spike potentials, the FN equations are a well-known non-linear ODE system, and we consider the mixed-effects version: 
\begin{equation}
\begin{cases}
V_j'(t) &= \theta_{j3}(V_j(t)-\frac{V^3_j(t)}{3}+R_j(t))\\
R_j'(t) &= -\frac{1}{\theta_{j3}}(V_j(t)-\theta_{j1}+\theta_{j2}R_j(t)),
\label{eqn:fn_model}
\end{cases}
\end{equation}
where for the $j$-th subject (or replicate), $V_j$ represents voltage of the neuron membrane potential, $R_j$ is the recovery variable from neuron currents, and $\boldsymbol{\theta}_j = (\theta_{j1}, \theta_{j2}, \theta_{j3})$ is the vector of subject-specific ODE parameters. 
As in  \citet{yang2021inference}, we set the true values of the fixed-effects parameters to be $\eta_1 = \eta_2 = 0.2$ and $\eta_3 = 3$. We generate  $\boldsymbol{\theta}_j =  (\eta_1,\eta_2,\eta_3)^\top + (b_{j1},b_{j2},b_{j3})^\top$ with $(b_{j1},b_{j2},b_{j3})^\top\sim N(\mathbf{0}, \boldsymbol{\Sigma}_{\boldsymbol{b}})$ for 25 subjects where 
$\boldsymbol{\Sigma}_{\boldsymbol{b}}=\begin{bsmallmatrix}
    0.0025 & 0.0025 & 0.03\\
     0.0025&  0.0025 & 0.03\\
     0.03 & 0.03 & 0.36
\end{bsmallmatrix}$. The initial conditions $V_j(0)$ and $R_j(0)$ are assumed to independently follow the normal distributions $V_j(0)\sim N(-1, 0.1^2)$ and $R_j(0)\sim N(1,0.1^2)$. To create the simulation data, we first use a numerical solver to generate the true trajectories for each subject, and then take 41 equally-spaced observations over the time interval $[0, 20]$ with noise SDs of $\sigma_{V} = \sigma_{R} =0.1$. We repeat this procedure to create 100 simulated datasets.
Implementation details for MAGI-ME on this example are provided in Section C.1 of the Appendix.

Table \ref{tab:param_tab_2} summarizes the average parameter estimates and corresponding standard deviations across 100 simulated data sets. MAGI-ME recovers the system well, as evidenced by the low standard deviation and bias. The fixed-effects parameter $\eta_2$ tends to be slightly overestimated, which in turn affects the inference of the corresponding entries in the covariance matrix $\boldsymbol{\Sigma}_{\boldsymbol{b}}(1,2), \boldsymbol{\Sigma}_{\boldsymbol{b}}(2,2)$, and $\boldsymbol{\Sigma}_{\boldsymbol{b}}(2,3)$. One possible explanation is that $R_j(t)$ may oscillate within a narrow range and the term $\theta_{j2}R_j(t)$ in \eqref{eqn:fn_model} stays numerically small, even with variation in $\theta_{j2}$. This makes the ODE parameters associated with $\theta_{j2}$ more difficult to estimate. Figure S2 in Section C.2 of the Appendix shows MAGI-ME's capability to recover the true underlying trajectories and provide reasonable 95$\%$ credible intervals. 
    \begin{table}[hbt!]
    \caption{Average parameter estimates using MAGI-ME (with standard deviation after $\pm$ sign) for the mixed-effects FN model across 100 simulated data sets. $\boldsymbol{\Sigma}_{\boldsymbol{b}}(a,b)$ denotes the $(a,b)$-th entry of $\boldsymbol{\Sigma}_{\boldsymbol{b}}$.}
    \resizebox{\textwidth}{!}{
\begin{tabular}{@{}clllclllclll@{}}
\toprule
Parameter  & \multicolumn{1}{c}{Truth} & \multicolumn{1}{c}{Estimate} 
           & Parameter & \multicolumn{1}{c}{Truth} & \multicolumn{1}{c}{Estimate}  
           & Parameter & \multicolumn{1}{c}{Truth} & \multicolumn{1}{c}{Estimate}  \\ \midrule
$\eta_1$   & 0.2       & 0.1991 $\pm$ 0.0106 
           & $\boldsymbol{\Sigma}_{\boldsymbol{b}}(1,1)$ & 0.0025 & 0.0029 $\pm$ 0.0006 
           & $\boldsymbol{\Sigma}_{\boldsymbol{b}}(2,2)$ & 0.0025 & 0.0062 $\pm$ 0.0014 \\
$\eta_2$   & 0.2       & 0.2368 $\pm$ 0.0182 
           & $\boldsymbol{\Sigma}_{\boldsymbol{b}}(1,2)$ & 0.0025 & 0.0037 $\pm$ 0.0009 
           & $\boldsymbol{\Sigma}_{\boldsymbol{b}}(2,3)$ & 0.03   & 0.0416 $\pm$ 0.0094 \\
$\eta_3$   & 3         & 2.9020 $\pm$ 0.1170 
           & $\boldsymbol{\Sigma}_{\boldsymbol{b}}(1,3)$ & 0.03   & 0.0310 $\pm$ 0.0067 
           & $\boldsymbol{\Sigma}_{\boldsymbol{b}}(3,3)$ & 0.36   & 0.3378 $\pm$ 0.0735 \\
\bottomrule
\end{tabular}
}
\label{tab:param_tab_2}
\end{table}

\section{Application to Pharmacokinetics}
\label{sec:application_chap_3}

Pharmacokinetics studies how the body is affected by a drug as it is absorbed, distributed, metabolized, and eventually eliminated \citep{goodman1996goodman}. An important research objective is to optimize therapeutic regimens based on the time-course concentration data from a treatment group of subjects. 
Drug exposure is typically assessed via key pharmacokinetic measures that include $C_{\max}$ (peak plasma concentration), $C_{\min}$ (trough plasma concentration), and AUC (area under the concentration-time curve). $C_{\max}$ levels above a threshold
may increase the risk of severe side effects, while drug efficacy may require maintaining $C_{\min}$ above a therapeutic threshold.
$\mathrm{AUC}$ quantifies the total post-dose drug exposure.
Clinical studies have focused on developing simplified dosing regimens that aim to reduce drug dosage while preserving antiviral efficacy \citep[e.g.,][]{wasmuth2004comparison}.

For an illustrative analysis using a publicly-available dataset, we consider the combination therapy of indinavir (IDV) and ritonavir (RTV) for HIV treatment first reported in \citet{wasmuth2004comparison}. The two treatment groups were prescribed different dosages: 400/100 mg IDV/RTV combination (`Treatment I', which enrolled 16 healthy participants) and 600/100 mg IDV/RTV combination (`Treatment II', which enrolled 15 healthy participants). 
The antiviral efficacy and risk of adverse effects are influenced by the plasma concentration of IDV: $C_{\max}>$ 8 mg/L is associated with side effects, while $C_{\min}<$ 0.1 mg/L may lead to a higher risk of viral replication.
IDV serum concentrations were collected at 0, 0.5, 1.0, 2.0, 2.5, 3.0, 4.0, 5.0, 6.0, 8.0, 10.0, and 12.0 hours after participants had taken the dosage twice daily for two weeks. In such studies, some measurements can be missing; e.g., the observation at $t=12$ is missing for the first subject under Treatment I. 

In the 12-hour post-dose period, the lowest observed concentration value
from both treatment groups remained above the threshold value, suggesting that drug effects may be sustained beyond 12 hours. To better characterize this terminal elimination phase, it may be necessary to extend the observation window or make predictions from the data. For instance, \citet{hsu1998pharmacokinetic} collected post-dose IDV plasma concentration over an 18-hour period. \citet{wang2014estimating} also estimated mixed-effects parameters and pharmacokinetic measures using the \citet{wasmuth2004comparison} data; however, that analysis had some limitations: (i) it lacked uncertainty quantification for the subject-specific pharmacokinetic measures and concentration-time curve, (ii) it could not predict (and quantify uncertainty of) the concentration-time curve beyond the 12-hour window, (iii) its limited ability to handle  missing data. In this section, we demonstrate how MAGI-ME addresses these limitations, by providing subject-specific uncertainty quantification of pharmacokinetic measures and predicted concentrations up to 18 hours post-dose.

The focus of our analysis is to estimate the fixed-effects parameters $Ka,Cl$, random-effects parameters $\sigma_{Ka},\sigma_{Cl}$, and pharmacokinetic measures $C_{\max},C_{\min},$ $\mathrm{AUC}$ for the two treatment groups using \eqref{eqn:pk/pd_model}. We  describe how to fit the model in \eqref{eqn:pk/pd_model} using the full set of observations from the two treatment groups in Section D.1 of the Appendix.
The estimated mixed-effects ODE parameters with 95$\%$ credible intervals obtained by MAGI-ME are summarized in Table \ref{tab:est_param_application}. The estimated population-level IDV elimination rates ($Ke$) under Treatment I and Treatment II are similar, with substantial overlap in their 95$\%$ credible intervals. This suggests that there is no strong evidence for a difference in elimination rates between the two treatment groups. In contrast, the estimated population-level IDV absorption rate ($Ka$) and clearance rate ($Cl$) under Treatment II are approximately 71$\%$ and 80$\%$, respectively, of those under Treatment I;  the associated posterior probabilities that $Ka$ and $Cl$ 
under Treatment I exceed those 
under Treatment II are 0.87 and 0.93, respectively.

Table \ref{tab:pharm_parameters} summarizes the estimated subject-specific pharmacokinetic measures and corresponding 95$\%$ credible intervals. On average, the estimated $C_{\max}$ under Treatment II is 1.6 times higher than under Treatment I.
Notably, for all subjects in Treatment I, both the estimated $C_{\max}$ and their 95$\%$ upper bounds remain below the toxicity threshold of 8 mg/L. In contrast, two subjects in Treatment II exhibit estimated $C_{\max}$ exceeding this threshold. This indicates that the higher IDV dosage of 600 mg may increase the risk of adverse side effects.
The average $C_{\min}$ estimate under Treatment II is 1.52 times higher than that under Treatment I. In particular, one subject in Treatment I has a $C_{\min}$ below the threshold value of 0.1, and 10 out of 16 subjects have 95$\%$ credible interval lower bounds falling below 0.1. In contrast, none of the subjects receiving Treatment II have $C_{\min}$ below the threshold value.
This indicates that the lower IDV dosage may potentially increase the risk of viral replication and resistance. Taken together, as some subjects treated with higher dosage may experience adverse effects, while others treated with reduced dosage may not achieve a therapeutic response, these results suggest a need for carefully individualized dosing regimens. 

Figure S3 in Section D.2 of the Appendix visualizes the inferred trajectories and associated 95$\%$ credible intervals for each individual subject using MAGI-ME, which appear to be reasonable fits.
Based on these trajectories, the estimated mean AUC in Treatment II was 82$\%$ higher than in Treatment I with limited overlap in subject-level 95$\%$ credible intervals, which suggests that the difference between an IDV dose of 600 mg and 400 mg is large enough to significantly change the total drug exposure.  

\begin{table}[hbt!]
\caption{Estimated fixed-effects parameters using MAGI-ME for the pharmacokinetic mixed-effects ODE model with 95$\%$ credible intervals, based on the observed IDV concentration from the two treatment groups.}
\resizebox{\textwidth}{!}{
\begin{tabular}{@{}ccccccccccccc@{}}
\toprule
 Parameter & \multicolumn{2}{c}{$Ke$}            & \multicolumn{2}{c}{$Ka$}            & \multicolumn{2}{c}{$Cl$}                & \multicolumn{2}{c}{$\sigma_{Ka}$}   & \multicolumn{2}{c}{$\sigma_{Cl}$}     \\ \midrule
Treatment          & I                & II               & I                & II               & I                  & II                 & I                & II               & I                & II                       \\
Estimate                & 0.30           & 0.27           & 1.00          & 0.71           & 22.45            & 18.02            & 0.50           & 0.31           & 5.84          & 4.22                  \\
95 $\%$ CI              & (0.26, 0.34) & (0.23, 0.32) & (0.74, 1.34) & (0.54, 0.95) & (19.36, 26.04) & (15.71, 20.67) & (0.30, 0.83) & (0.18, 0.55) & (3.61, 9.47) & (2.60, 6.85)  \\ \bottomrule
\end{tabular}}
\label{tab:est_param_application}
\end{table}

\begin{table}[htbp]
\caption{Estimated pharmacokinetic measures ($C_{\max}$, $C_{\min}$, and AUC) using MAGI-ME with 95\% credible intervals for each subject based on the observations from the two treatment groups.  }
\resizebox{\textwidth}{!}{%
\begin{tabular}{@{}ccccc|ccc|ccc|ccc@{}}
\toprule
Treatment            & Subject   & \multicolumn{3}{c|}{1}                        & \multicolumn{3}{c|}{2}                        & \multicolumn{3}{c|}{3}                        & \multicolumn{3}{c}{4}                        \\ \midrule
\multirow{15}{*}{I}  & Parameter & $C_{\max}$   & $C_{\min}$    & $\mathrm{AUC}$          & $C_{\max}$   & $C_{\min}$    & $\mathrm{AUC}$          & $C_{\max}$   & $C_{\min}$    & $\mathrm{AUC}$          & $C_{\max}$   & $C_{\min}$   & $\mathrm{AUC}$          \\
                     & Estimate  & 2.74         & 0.09          & 11.48          & 4.89         & 0.23          & 23.23          & 4.08         & 0.22          & 20.94          & 4.01         & 0.43         & 25.81          \\
                     & 95$\%$ CI & (2.25, 3.23) & (0.00, 0.56) & (7.30, 16.74)  & (4.41, 5.36) & (0.00, 0.48) & (19.03, 27.44) & (3.64, 4.53) & (0.00, 0.46) & (16.84, 25.06) & (3.60, 4.41) & (0.27, 0.59) & (21.49, 30.12) \\ \cmidrule(l){2-14} 
                     & Subject   & \multicolumn{3}{c|}{5}                        & \multicolumn{3}{c|}{6}                        & \multicolumn{3}{c|}{7}                        & \multicolumn{3}{c}{8}                        \\ \cmidrule(l){2-14} 
                     & Parameter & $C_{\max}$   & $C_{\min}$    & $\mathrm{AUC}$          & $C_{\max}$   & $C_{\min}$    & $\mathrm{AUC}$          & $C_{\max}$   & $C_{\min}$    & $\mathrm{AUC}$          & $C_{\max}$   & $C_{\min}$   & $\mathrm{AUC}$          \\
                     & Estimate  & 2.56         & 0.19          & 14.27          & 2.07         & 0.33          & 14.38          & 3.07         & 0.26          & 18.03          & 2.90         & 0.56         & 21.06          \\
                     & 95$\%$ CI & (2.18, 2.95) & (0.09, 0.28)  & (11.04, 17.54) & (1.70, 2.45) & (0.14, 0.52)  & (10.29, 18.60) & (2.66, 3.47) & (0.10, 0.42)  & (13.99, 22.12) & (2.51, 3.28) & (0.31, 0.80) & (16.65, 25.59) \\ \cmidrule(l){2-14} 
                     & Subject   & \multicolumn{3}{c|}{9}                        & \multicolumn{3}{c|}{10}                       & \multicolumn{3}{c|}{11}                       & \multicolumn{3}{c}{12}                       \\ \cmidrule(l){2-14} 
                     & Parameter & $C_{\max}$   & $C_{\min}$    & $\mathrm{AUC}$          & $C_{\max}$   & $C_{\min}$    & $\mathrm{AUC}$          & $C_{\max}$   & $C_{\min}$    & $\mathrm{AUC}$          & $C_{\max}$   & $C_{\min}$   & $\mathrm{AUC}$          \\
                     & Estimate  & 3.24         & 0.28          & 19.30          & 3.97         & 0.21          & 19.45          & 3.08         & 0.20          & 16.52          & 3.47         & 0.51         & 23.96          \\
                     & 95$\%$ CI & (2.82, 3.65) & (0.07, 0.49)  & (14.95, 23.68) & (3.53, 4.41) & (0.03, 0.38)  & (15.84, 23.07) & (2.67, 3.49) & (0.05, 0.36)  & (12.97, 20.14) & (3.04, 3.91) & (0.13, 0.90) & (18.61, 29.37) \\ \cmidrule(l){2-14} 
                     & Subject   & \multicolumn{3}{c|}{13}                       & \multicolumn{3}{c|}{14}                       & \multicolumn{3}{c|}{15}                       & \multicolumn{3}{c}{16}                       \\ \cmidrule(l){2-14} 
                     & Parameter & $C_{\max}$   & $C_{\min}$    & $\mathrm{AUC}$          & $C_{\max}$   & $C_{\min}$    & $\mathrm{AUC}$          & $C_{\max}$   & $C_{\min}$    & $\mathrm{AUC}$          & $C_{\max}$   & $C_{\min}$   & $\mathrm{AUC}$          \\
                     & Estimate  & 2.68         & 0.26          & 16.39          & 2.66         & 0.13          & 12.55          & 4.14         & 0.24          & 20.79          & 4.73         & 0.53         & 30.38          \\
                     & 95$\%$ CI & (2.30, 3.07) & (0.13, 0.39)  & (12.63, 20.23) & (2.23, 3.10) & (0.00, 0.33) & (9.12, 16.07)  & (3.68, 4.59) & (0.02, 0.46)  & (16.73, 24.85) & (4.32, 5.15) & (0.29, 0.77) & (25.67, 35.15) \\ \midrule
                     & Subject   & \multicolumn{3}{c|}{1}                        & \multicolumn{3}{c|}{2}                        & \multicolumn{3}{c|}{3}                        & \multicolumn{3}{c}{4}                        \\ \cmidrule(l){2-14} 
\multirow{15}{*}{II} & Parameter & $C_{\max}$   & $C_{\min}$    & $\mathrm{AUC}$          & $C_{\max}$   & $C_{\min}$    & $\mathrm{AUC}$          & $C_{\max}$   & $C_{\min}$    & $\mathrm{AUC}$          & $C_{\max}$   & $C_{\min}$   & $\mathrm{AUC}$          \\
                     & Estimate  & 4.79         & 0.20          & 22.05          & 8.20         & 1.10          & 55.28          & 4.48         & 0.67          & 31.12          & 5.47         & 1.48         & 43.88          \\
                     & 95$\%$ CI & (4.00, 5.57) & (0.00, 0.92) & (14.67, 30.74) & (7.58, 8.83) & (0.77, 1.43)  & (48.12, 62.43) & (3.92, 5.04) & (0.40, 0.94)  & (24.95, 37.30) & (4.89, 6.04) & (0.99, 1.97) & (36.67, 51.12) \\ \cmidrule(l){2-14} 
                     & Subject   & \multicolumn{3}{c|}{5}                        & \multicolumn{3}{c|}{6}                        & \multicolumn{3}{c|}{7}                        & \multicolumn{3}{c}{8}                        \\ \cmidrule(l){2-14} 
                     & Parameter & $C_{\max}$   & $C_{\min}$    & $\mathrm{AUC}$          & $C_{\max}$   & $C_{\min}$    & $\mathrm{AUC}$          & $C_{\max}$   & $C_{\min}$    & $\mathrm{AUC}$          & $C_{\max}$   & $C_{\min}$   & $\mathrm{AUC}$          \\
                     & Estimate  & 4.20         & 0.64          & 28.73          & 4.59         & 0.49          & 28.20          & 6.08         & 0.45          & 33.36          & 4.93         & 0.99         & 36.41          \\
                     & 95$\%$ CI & (3.64, 4.77) & (0.37, 0.92)  & (22.58, 35.04) & (4.00, 5.17) & (0.24, 0.73)  & (22.46, 34.04) & (5.43, 6.74) & (0.04, 0.85)  & (26.85, 39.88) & (4.36, 5.51) & (0.63, 1.35) & (29.79, 43.17) \\ \cmidrule(l){2-14} 
                     & Subject   & \multicolumn{3}{c|}{9}                        & \multicolumn{3}{c|}{10}                       & \multicolumn{3}{c|}{11}                       & \multicolumn{3}{c}{12}                       \\ \cmidrule(l){2-14} 
                     & Parameter & $C_{\max}$   & $C_{\min}$    & $\mathrm{AUC}$          & $C_{\max}$   & $C_{\min}$    & $\mathrm{AUC}$          & $C_{\max}$   & $C_{\min}$    & $\mathrm{AUC}$          & $C_{\max}$   & $C_{\min}$   & $\mathrm{AUC}$          \\
                     & Estimate  & 6.93         & 0.64          & 41.41          & 4.01         & 0.54          & 26.30          & 5.57         & 0.78          & 37.85          & 5.12         & 0.49         & 30.46          \\
                     & 95$\%$ CI & (6.32, 7.55) & (0.41, 0.85)  & (35.57, 47.26) & (3.44, 4.58) & (0.27, 0.82)  & (20.16, 32.64) & (4.98, 6.16) & (0.51, 1.06)  & (31.42, 44.29) & (4.52, 5.72) & (0.21, 0.76) & (24.50, 36.47) \\ \cmidrule(l){2-14} 
                     & Subject   & \multicolumn{3}{c|}{13}                       & \multicolumn{3}{c|}{14}                       & \multicolumn{3}{c|}{15}                       &              &              &                \\ \cmidrule(l){2-14} 
                     & Parameter & $C_{\max}$   & $C_{\min}$    & $\mathrm{AUC}$          & $C_{\max}$   & $C_{\min}$    & $\mathrm{AUC}$          & $C_{\max}$   & $C_{\min}$    & $\mathrm{AUC}$          &              &              &                \\
                     & Estimate  & 3.62         & 0.31          & 20.56          & 5.35         & 1.47          & 44.14          & 8.04         & 0.69          & 46.27          &              &              &                \\
                     & 95$\%$ CI & (3.02, 4.23) & (0.00, 0.78) & (14.05, 27.39) & (4.77, 5.93) & (0.91, 2.04)  & (36.67, 51.61) & (7.38, 8.71) & (0.31, 1.07)  & (39.51, 53.05) &              &              &                \\ \bottomrule
\end{tabular}}
\footnotesize
Notes: \\
For several subjects, the lower bound of the normal-based 95$\%$ credible interval for $C_{\min}$ is truncated to zero to ensure interpretability in the context of IDV concentration.
\label{tab:pharm_parameters}
\end{table}

\begin{figure}[htbp]

    \centering
    \begin{subfigure}{\linewidth}
        \centering
        \includegraphics[width=\linewidth]{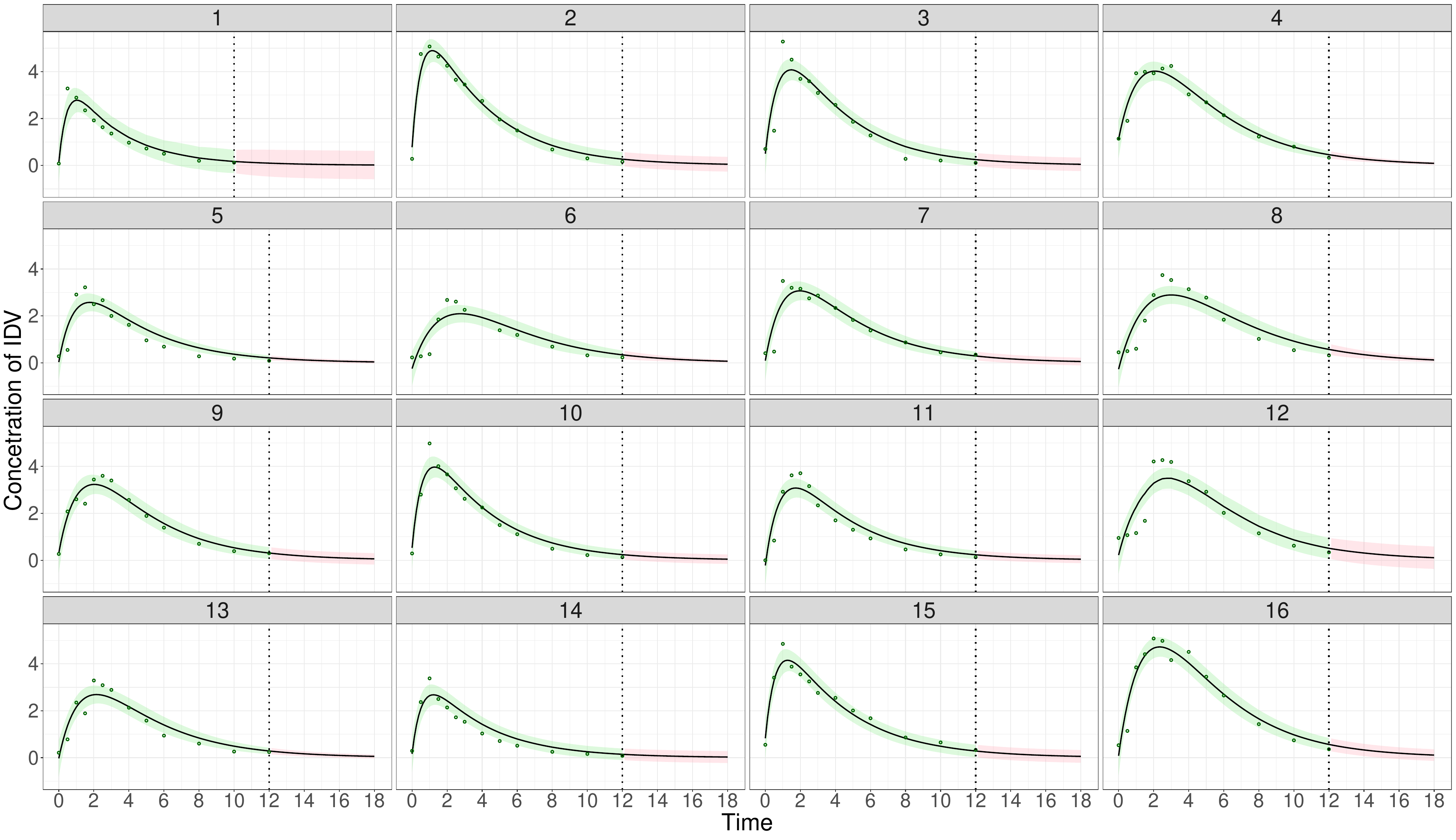}
        \caption{Inferred trajectories for subjects receiving Treatment I.}
        
    \end{subfigure}
    \begin{subfigure}{\linewidth}
        \centering
        \includegraphics[width=\linewidth]{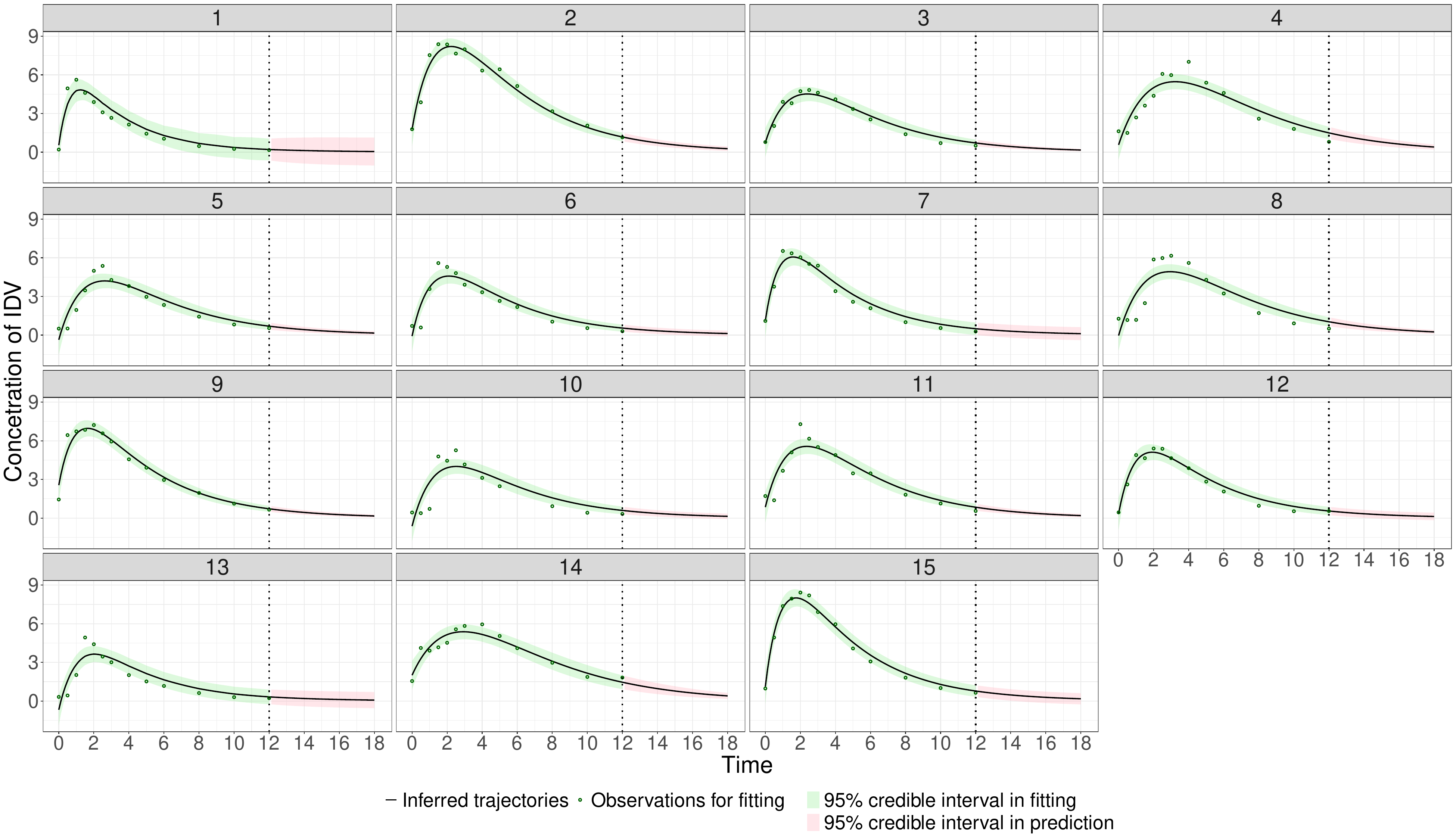}
        \caption{Inferred trajectories for subjects receiving Treatment II.}

    \end{subfigure}

    \caption{Inferred trajectories for the pharmacokinetic mixed-effects model based on the IDV concentration data. The vertical dotted line separates the fitting and prediction periods. The green dots are the observed data.
    The black line represents the inferred trajectory, with the green shaded area representing the 95$\%$ pointwise credible interval for the fitting period, and the pink shaded area representing the 95$\%$ pointwise credible interval for the prediction period.}
    \label{fig:data_pred}
\end{figure}

From the fit to the 12-hour IDV concentration data, subjects receiving Treatment I generally exhibit IDV trough concentration that is closer to the threshold value compared to those receiving Treatment II. We further evaluate the persistence of drug exposure and efficacy by predicting IDV concentration for the subsequent 6 hours in both treatment groups. To set up MAGI-ME for prediction, we extend the discretization set to $\boldsymbol{I}_j = \{0,0.125,\cdots, 18\}$ for each subject in the two treatment groups to cover both the fitting and prediction periods. The model is fitted using the same data (i.e., up to 12 hours of observations only) and implementation details as before. Predictions for the subsequent 6-hour window are then provided by MAGI-ME's inferred trajectory for $12 < t \le 18$. By applying a normal approximation to  $C_{\min}|(\boldsymbol{W}_{\boldsymbol{I}} = 0, \boldsymbol{y}(\boldsymbol{\gamma}))$, based on the normality assumption in \eqref{eqn:normality_assumption}, we can estimate the posterior probability that the trough concentration falls below the threshold value after 18 hours. 
Table \ref{tab:pred_param} presents, for each subject, the predicted trough concentration at 18 hours and the corresponding probability of falling below the threshold value. 
Subjects under Treatment I are more likely to exhibit an IDV trough concentration below the threshold value compared to those receiving Treatment II, as visualized in Figure \ref{fig:data_pred}: by 18 hours post-dose, 13 out of 16 subjects under Treatment I have predicted concentrations below the threshold, in contrast to 2 out of 15 subjects under Treatment II.

\begin{table}[hbt!]
\caption{Predicted trough concentration ($C_{\min}$) of IDV after 18 hours based on observations from 0 to 12 hours in the two treatment groups. The bottom row reports the probability that $C_{\min}$ falls below the threshold value of 0.1 by 18 hours post-dose. }
\resizebox{\textwidth}{!}{
\begin{tabular}{@{}cccccccccc@{}}
\toprule
\multicolumn{1}{l}{Treatment} & Subject           & 1             & 2             & 3             & 4             & 5             & 6             & 7             & 8             \\ \midrule
\multirow{7}{*}{I}            & $C_{\min}$        & 0.02          & 0.05          & 0.05          & 0.09          & 0.04          & 0.07          & 0.06          & 0.12          \\
                              & $P(C_{\min}<0.1)$ & 0.60          & 0.62          & 0.64          & 0.62          & 0.94          & 0.83          & 0.70          & 0.31          \\ \cmidrule(l){2-10} 
                              & Subject           & 9             & 10            & 11            & 12            & 13            & 14            & 15            & 16            \\ \cmidrule(l){2-10} 
                              & $C_{\min}$        & 0.06          & 0.05          & 0.05          & 0.11          & 0.06          & 0.03          & 0.06          & 0.11          \\
                              & $P(C_{\min}<0.1)$ & 0.63          & 0.70          & 0.74          & 0.49          & 0.84          & 0.71          & 0.63          & 0.47          \\ \midrule
\multicolumn{1}{l}{Treatment} & Subject           & 1             & 2             & 3             & 4             & 5             & 6             & 7             & 8             \\ \midrule
\multirow{7}{*}{II}           & $C_{\min}$        & 0.04          & 0.27          & 0.16          & 0.40          & 0.16          & 0.12          & 0.11          & 0.25          \\
                       
                              & $P(C_{\min}<0.1)$ & 0.54          & 0.03          & 0.10          & 0.00          & 0.23          & 0.42          & 0.49          & 0.03          \\ \cmidrule(l){2-10} 
                              & Subject           & 9             & 10            & 11            & 12            & 13            & 14            & 15            &               \\ \cmidrule(l){2-10} 
                              & $C_{\min}$        & 0.16          & 0.13          & 0.19          & 0.12          & 0.07          & 0.40          & 0.17          &               \\
                             
                              & $P(C_{\min}<0.1)$ & 0.15          & 0.38          & 0.08          & 0.44          & 0.53          & 0.01          & 0.37          &               \\ \bottomrule
\end{tabular}}

\label{tab:pred_param}
\end{table}
\section{Discussion}
\label{sec:discussion_chap_3}
In this paper, we presented MAGI-ME for inference of general mixed-effects ODE models.
MAGI-ME conducts inference using a normal approximation to the posterior distribution.
A numerical study using a homogeneous ODE model indicates that MAGI-ME achieves parameter inference and trajectory recovery comparable to both the  \texttt{nlme()} function and MCMC sampling from the full MAGI posterior. Simulation results indicate that MAGI-ME well-recovers the parameters and trajectories in a non-homogeneous ODE model compared to other representative Bayesian methods, and in a multi-component FN mixed-effects model that demonstrates its capability to estimate more complex systems; neither of these models have analytic solutions. Finally, we fit a pharmacokinetic ODE model using time-course concentration data as a practical application. MAGI-ME readily provides uncertainty quantification of pharmacokinetic measures and predicted concentration curves, thereby offering a more comprehensive interpretation of clinical data.

We outline some directions for future research. First, we could relax the assumption of multivariate normality for  measurement errors and random effects. The limitations of such an assumption, and the potential violations of normality among subject-specific ODE parameters in practice, have been discussed in \citet{liu2019bayesian}. MAGI-ME provides a flexible framework to accommodate non-normality, as the measurement error and random effects distributions can be easily changed in the Bayesian hierarchical model.  
Second, another interesting area is the development of ODE model selection tools. 
In pharmacokinetic studies, when multiple ODE models can be used for the same time-course concentration data, it is important to identify the most appropriate model among all the possible ones \citep{mcdonough2023using}. Traditional selection criteria, such as AIC and BIC, have been commonly used \citep{wu1999population,liang2010estimation}. However, AIC and BIC rely solely on point estimates and do not account for uncertainty in parameter estimation \citep{gelman2014understanding}. The challenge of evaluating many candidate models can be especially pronounced for the higher dimensionality associated with mixed-effects ODEs. 
\bibliography{ref}
\appendix
\renewcommand{\thesection}{Appendix \Alph{section}}
\renewcommand{\thefigure}{Figure S\arabic{figure}}
\captionsetup[figure]{labelformat=empty} 
\renewcommand{\thetable}{Table S\arabic{table}}
\captionsetup[table]{labelformat=empty} 

\setcounter{table}{0}
\setcounter{figure}{0}
\setcounter{theorem}{0}

\newpage
\clearpage
\begin{center}
  \LARGE Supplementary Material for Manifold-Constrained Gaussian Processes for Inference of Mixed-effects Ordinary Differential Equations with Application to Pharmacokinetics by Yuxuan Zhao and Samuel W.K. Wong
\end{center}
\title{}
\maketitle
\section{Posterior Inference Details}
In this section, we provide the details of posterior inference under the MAGI-ME framework. We first present the full expression of the posterior distribution. Then, we detail the procedures for the nested levels of optimization. For uncertainty quantification, we derive the standard error of the random-effects parameter $\boldsymbol{u}$ using the Delta method of \citet{kass1989approximate}. Last, we conduct a numerical validation study to support the reliability of our inference results.

\subsection{Detailed Expression of  Posterior Distribution in MAGI-ME}
This section gives the details of the posterior distribution of MAGI-ME. Factorizing the left hand side of (6) in the main text yields

$$
\begin{aligned}
&p\left(\boldsymbol{\eta},\boldsymbol{\Sigma}_{\boldsymbol{b}},\boldsymbol{b},  \boldsymbol{x}(\boldsymbol{I})|\boldsymbol{W}_{\boldsymbol{I}} = \boldsymbol{0}, \boldsymbol{y}(\boldsymbol{\gamma})\right) \propto p\left(\boldsymbol{\eta},\boldsymbol{\Sigma}_{\boldsymbol{b}},\boldsymbol{b},  \boldsymbol{x}(\boldsymbol{I}), \boldsymbol{W}_{\boldsymbol{I}} = \boldsymbol{0}, \boldsymbol{y}(\boldsymbol{\gamma})\right)
\\ & = \underbrace{\pi(\boldsymbol{\eta})}_{(1)} \times \underbrace{\pi(\boldsymbol{\Sigma}_{\boldsymbol{b}})}_{(2)}\times\underbrace{p(\boldsymbol{b}|\boldsymbol{\Sigma}_{\boldsymbol{b}})}_{(3)}\times \underbrace{p(\boldsymbol{x}(\boldsymbol{I})|\boldsymbol{b},\boldsymbol{\eta},\boldsymbol{\Sigma}_{\boldsymbol{b}})}_{(4)}
\times \underbrace{p(\boldsymbol{y}(\boldsymbol{\gamma})|\boldsymbol{x}(\boldsymbol{I}),\boldsymbol{b},\boldsymbol{\eta},\boldsymbol{\Sigma}_{\boldsymbol{b}})}_{(5)}\\
&\times \underbrace{p(\boldsymbol{W}_{\boldsymbol{I}} = \boldsymbol{0}|\boldsymbol{y}(\boldsymbol{\gamma}),\boldsymbol{x}(\boldsymbol{I}),\boldsymbol{b},\boldsymbol{\eta},\boldsymbol{\Sigma}_{\boldsymbol{b}})}_{(6)}.
\end{aligned}
$$
The fourth term can be simplified as $p(\boldsymbol{x}(\boldsymbol{I}))$ due to the prior independence between $\boldsymbol{x}(\boldsymbol{I})$ and $\boldsymbol{b}$, $\boldsymbol{\eta}$, $\boldsymbol{\Sigma}_{\boldsymbol{b}}$. 
The fifth term corresponds to the likelihood of the observations and likewise does not depend on $\boldsymbol{b}$, $\boldsymbol{\eta}$, $\boldsymbol{\Sigma}_{\boldsymbol{b}}$, so can be written as $p(\boldsymbol{y}(\boldsymbol{\gamma})|\boldsymbol{x}(\boldsymbol{I}))$. To simplify the sixth term, we substitute the definition of $\boldsymbol{W}_{\boldsymbol{I}}=\boldsymbol{0}$ and note that the resulting GP derivative 
$\boldsymbol{x}'(\boldsymbol{I})|\boldsymbol{x}(\boldsymbol{I})$ is conditionally independent of $\boldsymbol{b}$, $\boldsymbol{\eta}$, $\boldsymbol{\Sigma}_{\boldsymbol{b}}$, and the observations $\boldsymbol{y}(\boldsymbol{\gamma})$:
$$\begin{aligned}
   & p(\boldsymbol{W}_{\boldsymbol{I}} = \boldsymbol{0}|\boldsymbol{y}(\boldsymbol{\gamma}),\boldsymbol{x}(\boldsymbol{I}),\boldsymbol{b},\boldsymbol{\eta},\boldsymbol{\Sigma}_{\boldsymbol{b}}) 
=p(\boldsymbol{x}'(\boldsymbol{I}) = \mathbf{f}(\boldsymbol{x}(\boldsymbol{I}), \boldsymbol{\eta}, \boldsymbol{b},\boldsymbol{I})|\boldsymbol{y}(\boldsymbol{\gamma}),\boldsymbol{x}(\boldsymbol{I}),\boldsymbol{b},\boldsymbol{\eta},\boldsymbol{\Sigma}_{\boldsymbol{b}})\\
   &=p(\boldsymbol{x}'(\boldsymbol{I}) = \mathbf{f}(\boldsymbol{x}(\boldsymbol{I}), \boldsymbol{\eta}, \boldsymbol{b},\boldsymbol{I})|\boldsymbol{x}(\boldsymbol{I})).
\end{aligned}$$
 Therefore, we  obtain 
\begin{equation*}
\begin{aligned}
&p\left(\boldsymbol{\eta},\boldsymbol{\Sigma}_{\boldsymbol{b}},\boldsymbol{b},  \boldsymbol{x}(\boldsymbol{I}), \boldsymbol{W}_{\boldsymbol{I}} = \boldsymbol{0}, \boldsymbol{y}(\boldsymbol{\gamma})\right)=\underbrace{\pi(\boldsymbol{\eta})}_{(1)}\times \underbrace{\pi(\boldsymbol{\Sigma}_{\boldsymbol{b}})}_{(2)}\times\underbrace{p(\boldsymbol{b}|\boldsymbol{\Sigma}_{\boldsymbol{b}})}_{(3)}\times \underbrace{p(\boldsymbol{x}(\boldsymbol{I}))}_{(4)}
\times \underbrace{p(\boldsymbol{y}(\boldsymbol{\gamma})|\boldsymbol{x}(\boldsymbol{I}))}_{(5)}\\
&\times \underbrace{p(\boldsymbol{x}'(\boldsymbol{I}) = \mathbf{f}(\boldsymbol{x}(\boldsymbol{I}), \boldsymbol{\eta}, \boldsymbol{b},\boldsymbol{I})|\boldsymbol{x}(\boldsymbol{I}))}_{(6)}.
\end{aligned}
\end{equation*}
The full expression of the computable posterior distribution is given by  $$
\begin{aligned}
  & p\left(\boldsymbol{\eta},\boldsymbol{\Sigma}_{\boldsymbol{b}},\boldsymbol{b},  \boldsymbol{x}(\boldsymbol{I}), \boldsymbol{W}_{\boldsymbol{I}} = \boldsymbol{0}, \boldsymbol{y}(\boldsymbol{\gamma})\right)
  =\underbrace{ \pi(\boldsymbol{\eta})}_{(1)}\times  \underbrace{\pi(\boldsymbol{\Sigma}_{\boldsymbol{b}})}_{(2)}\\
  &\times\underbrace{\exp\left[-\frac{1}{2}\sum_{j=1}^s |\boldsymbol{b}_j|\log (2\pi) + \log(\det (\boldsymbol{\Sigma}_{\boldsymbol{b}})) + \left\|\boldsymbol{b}_j\right\|^2_{\boldsymbol{\Sigma}^{-1}_{\boldsymbol{b}}}\right]}_{(3)}\\
  &\times  \exp \left\{-\frac{1}{2} \sum_{j=1}^s\sum_{i=1}^m \left[\underbrace{|\boldsymbol{I}_j| \log (2 \pi)+\log(\det (\boldsymbol{C}_{ij}))+\left\|\boldsymbol{x}_{ij}(\boldsymbol{I}_j)-\boldsymbol{\mu}_{ij}(\boldsymbol{I}_j)\right\|_{\boldsymbol{C}_{ij}^{-1}}^2}_{(4)} \right.\right.\\
    &\quad +\underbrace{N_{ij}\log \left(2 \pi \sigma_{ij}^2\right)+\left\|\boldsymbol{x}_{ij}\left(\boldsymbol{\gamma}_{ij}\right)-\boldsymbol{y}_{ij}\left(\boldsymbol{\gamma}_{ij}\right)\right\|_{\sigma_{ij}^{-2}}^2}_{(5)} \\
    &\left.\left.\quad +\underbrace{|\boldsymbol{I}_j| \log (2 \pi)+\log(\det (\boldsymbol{\zeta}_{ij}))+\left\|\mathbf{f}_{ij, \boldsymbol{I}_j}^{\boldsymbol{x}, \boldsymbol{\eta}, \boldsymbol{b}_j}-\boldsymbol{\mu}'_{ij}(\boldsymbol{I}_j)-\boldsymbol{m}_{ij}\left\{\boldsymbol{x}_{ij}(\boldsymbol{I}_j)-\boldsymbol{\mu}_{ij}(\boldsymbol{I}_j)\right\}\right\|_{\boldsymbol{\zeta}_{ij}^{-1}}^2}_{(6)} \right]\right\},\\
    \end{aligned}
$$
where $\|\boldsymbol{v}\|_A^2=\boldsymbol{v}^{\boldsymbol{\top}} A \boldsymbol{v}$,$\mathbf{f}_{ij, \boldsymbol{I}_j}^{\boldsymbol{x}. \boldsymbol{\eta},\boldsymbol{b}_j}$ represents the $i$-th component of $j-$th subject of $\mathbf{f}(\boldsymbol{x}_i(t),\boldsymbol{b}_{j},\boldsymbol{\eta}, \boldsymbol{I}_j)$, and $|\boldsymbol{I}_j|$ is the cardinality of $\boldsymbol{I}_j$. For each component $i$ in the $j-$th subject, $\sigma_{ij}$ denotes the noise level and $N_{ij}$ is the number of observations. The multivariate
normal covariance matrices are computed by 
$$\begin{cases}
\boldsymbol{C}_{ij} &= \mathcal{K}_{ij}(\boldsymbol{I}_j,\boldsymbol{I}_j)\\
\boldsymbol{m}_{ij} &={ }^{\prime} \mathcal{K}_{ij}(\boldsymbol{I}_j, \boldsymbol{I}_j) \mathcal{K}_{ij}(\boldsymbol{I}_j, \boldsymbol{I}_j)^{-1} \\
\boldsymbol{\zeta}_{ij} &=\mathcal{K}_{ij}^{\prime \prime}(\boldsymbol{I}_j, \boldsymbol{I}_j)-{ }^{\prime} \mathcal{K}_{ij}(\boldsymbol{I}_j, \boldsymbol{I}_j) \mathcal{K}_{ij}(\boldsymbol{I}_j, \boldsymbol{I}_j)^{-1} \mathcal{K}^{\prime}_{ij}(\boldsymbol{I}_j, \boldsymbol{I}_j)
\end{cases},$$ with
 ${ }^{\prime} \mathcal{K}_{ij}=\frac{\partial}{\partial s} \mathcal{K}_{ij}(s, t), \mathcal{K}^{\prime}_{ij}=\frac{\partial}{\partial t} \mathcal{K}_{ij}(s, t) \text {, and } \mathcal{K}^{\prime \prime}_{ij}=\frac{\partial^2}{\partial s \partial t} \mathcal{K}_{ij}(s, t)$.

\subsection{Nested Optimization Details}
In this section, we describe the details for the nested levels of optimization. 
The marginal posterior distribution of $\boldsymbol{\omega}$ can be obtained by integrating out $\boldsymbol{u}$ from the joint posterior, i.e.,
$p(\boldsymbol{\omega}|\boldsymbol{y}(\boldsymbol{\gamma}), \boldsymbol{W}_{\boldsymbol{I}}=0)
   = \int p(\boldsymbol{u},\boldsymbol{\omega}| \boldsymbol{W}_{\boldsymbol{I}} = \boldsymbol{0}, \boldsymbol{y}(\boldsymbol{\gamma})) d\boldsymbol{u}$. 
Maximizing this marginal posterior to obtain the fixed effects and covariance matrix of random effects is equivalent to maximizing \begin{equation}L(\boldsymbol{\omega}) =  \int \exp(-Q(\boldsymbol{u},\boldsymbol{\omega} )) d\boldsymbol{u},
\label{eqn:L_eta}
\end{equation}
where $ Q(\boldsymbol{u},\boldsymbol{\omega}) = -\log p(\boldsymbol{u},\boldsymbol{\omega}| \boldsymbol{W}_{\boldsymbol{I}} = \boldsymbol{0}, \boldsymbol{y}(\boldsymbol{\gamma}))$. This integral is intractable and we suggest using the Laplace approximation to approximate it. We fix $\boldsymbol{\omega}$ and take a second order Taylor expansion of $Q(\boldsymbol{u},\boldsymbol{\omega})$  around the mode $\hat{\boldsymbol{u}}(\boldsymbol{\omega})=\arg\min_{\boldsymbol{u}} Q(\boldsymbol{u},\boldsymbol{\omega})$ (which is the minimizer of $Q(\boldsymbol{u},\boldsymbol{\omega})$ with respect to $\boldsymbol{u}$), yielding
\begin{equation}
\begin{aligned}
&Q(\boldsymbol{u},\boldsymbol{\omega})\approx Q( \hat{\boldsymbol{u}}(\boldsymbol{\omega}), \boldsymbol{\omega})+\frac{\partial }{\partial \boldsymbol{u}}Q(\hat{\boldsymbol{u}}(\boldsymbol{\omega}), \boldsymbol{\omega})^\top (\boldsymbol{u}- \hat{\boldsymbol{u}}(\boldsymbol{\omega}))+\frac{1}{2}(\boldsymbol{u}- \hat{\boldsymbol{u}}(\boldsymbol{\omega}))^\top \boldsymbol{H}({\boldsymbol{\omega}})(\boldsymbol{u}- \hat{\boldsymbol{u}}(\boldsymbol{\omega})),
\end{aligned}
\label{eqn:taylor_expansion}
\end{equation}
where $\boldsymbol{H}(\boldsymbol{\omega})=\frac{\partial^2}{\partial \boldsymbol{u}\partial \boldsymbol{u}^\top}Q(\boldsymbol{u},\boldsymbol{\omega})|_{\boldsymbol{u} = \hat{\boldsymbol{u}}(\boldsymbol{\omega})}$ is the Hessian matrix of $Q(\boldsymbol{u},\boldsymbol{\omega})$ evaluated at $\hat{\boldsymbol{u}}(\boldsymbol{\omega})$. Noting that $\frac{\partial }{\partial \boldsymbol{u}}Q(\hat{\boldsymbol{u}}(\boldsymbol{\omega}), \boldsymbol{\omega}) = 0$  at the critical point $\hat{\boldsymbol{u}}(\boldsymbol{\omega})$,  the first-order term is zero. Writing the inverse of $\boldsymbol{H}(\boldsymbol{\omega})$ as $\boldsymbol{\Sigma}_{\boldsymbol{u}}(\boldsymbol{\omega})$ and plugging in \eqref{eqn:taylor_expansion} to \eqref{eqn:L_eta} yields the approximation $\tilde{L}(\boldsymbol{\omega})$ for the marginal posterior distribution of $\boldsymbol{\omega}$:
\begin{equation}
\begin{aligned}
    \tilde{L}(\boldsymbol{\omega})&= \int \exp\left\{-Q( \hat{\boldsymbol{u}}(\boldsymbol{\omega}), \boldsymbol{\omega}) - \frac{1}{2}(\boldsymbol{u}- \hat{\boldsymbol{u}}(\boldsymbol{\omega}))^\top \boldsymbol{H}({\boldsymbol{\omega}})(\boldsymbol{u}- \hat{\boldsymbol{u}}(\boldsymbol{\omega}))\right\}d\boldsymbol{u}\\
    &=\exp \left[-  Q( \hat{\boldsymbol{u}}(\boldsymbol{\omega}), \boldsymbol{\omega})+\frac{1}{2}\log\det(\boldsymbol{\Sigma}_{\boldsymbol{u}}(\boldsymbol{\omega}))\right](2\pi)^{|\boldsymbol{u}|/2}.
\end{aligned}
\label{eqn:marginal_eta}
\end{equation}
We obtain $\hat{\boldsymbol{\omega}}$ by minimizing $-\log \tilde{L}(\boldsymbol{\omega})$ using the BFGS algorithm, which also yields an estimate of its corresponding covariance matrix, i.e.,  $\hat{\boldsymbol{\Sigma}}_{\boldsymbol{\omega}}(\boldsymbol{\hat{\boldsymbol{\omega}}}) = -(\frac{\partial^2\tilde{L}(\boldsymbol{\omega})}{\partial\boldsymbol{\omega}\partial\boldsymbol{\omega}^\top})^{-1}|_{\boldsymbol{\omega}= \hat{\boldsymbol{\omega}}}$.

\subsection{Derivation of Standard Errors of Random-effects Parameters}
In this section, we derive the standard error of the random-effects parameter $\boldsymbol{u}$ using the Delta method of \citet{kass1989approximate} and construct the $(1-\alpha)$ approximate credible interval. 

We first derive the joint covariance matrix for $(\boldsymbol{u},\boldsymbol{\omega})^\top$. Under the approximate joint normality assumption for $(\boldsymbol{u},\boldsymbol{\omega})|(\boldsymbol{W}_{\boldsymbol{I}}=0, \boldsymbol{y}(\boldsymbol{\gamma}))$ in (7) of the main text, we can obtain the standard error for $\boldsymbol{u}$ by the Delta method  \citep{kass1989approximate} to account for the uncertainty in $\hat{\boldsymbol{\omega}}$, and derive approximate $(1-\alpha)$ credible intervals. 
By the law of total variance, we can decompose the variance of $(\boldsymbol{u},\boldsymbol{\omega})^\top$ into two terms:\begin{equation}
    \mathrm{Var}\left(\begin{bmatrix}
        \boldsymbol{u}\\
         \boldsymbol{\omega}
    \end{bmatrix}\right) = \underbrace{\mathrm{Var}\left[\mathrm{E}\left(\begin{bmatrix}
       \boldsymbol{u}\\
         \boldsymbol{\omega}
    \end{bmatrix}\bigg|\boldsymbol{\omega},\boldsymbol{W}_{\boldsymbol{I}}=0, \boldsymbol{y}(\boldsymbol{\gamma})\right)\right]}_{(1)} + \underbrace{\mathrm{E}\left[\mathrm{Var}\left(\begin{bmatrix}
       \boldsymbol{u}\\
         \boldsymbol{\omega}
    \end{bmatrix}\bigg|\boldsymbol{\omega},\boldsymbol{W}_{\boldsymbol{I}}=0, \boldsymbol{y}(\boldsymbol{\gamma})\right)\right]}_{(2)}.
    \label{eqn:decompose_variance}
\end{equation}
Noting that $\boldsymbol{u}|(\boldsymbol{\omega},\boldsymbol{W}_{\boldsymbol{I}}=0, \boldsymbol{y}(\boldsymbol{\gamma}))$ is also approximately normally distributed with mean $\hat{\boldsymbol{u}}(\boldsymbol{\omega})$ and covariance matrix $\boldsymbol{\Sigma}_{\boldsymbol{u}}(\boldsymbol{\omega})$, we can rewrite \eqref{eqn:decompose_variance} by $$
    \mathrm{Var}\left(\begin{bmatrix}
        \boldsymbol{u}\\
         \boldsymbol{\omega}
    \end{bmatrix}\right) = \underbrace{\mathrm{Var}\left(\begin{bmatrix}
       \hat{\boldsymbol{u}}(\boldsymbol{\omega})\\
         \boldsymbol{\omega}
    \end{bmatrix}\right)}_{(1)} + \underbrace{\mathrm{E}\left(\begin{bmatrix}
     \boldsymbol{\Sigma}_{\boldsymbol{u}}(\boldsymbol{\omega}) & 0 \\
      0 & 0
    \end{bmatrix}\right)}_{(2)}.
$$
Applying the standard Delta method to the first term gives 
$$
\mathrm{Var}\left(\begin{bmatrix}
       \hat{\boldsymbol{u}}(\boldsymbol{\omega})\\
         \boldsymbol{\omega}
    \end{bmatrix}\right) = \boldsymbol{J}^\top (\boldsymbol{\omega})\boldsymbol{\Sigma}_{\boldsymbol{\omega}}(\boldsymbol{\omega})\boldsymbol{J}(\boldsymbol{\omega}),
$$
where  $\boldsymbol{J} (\boldsymbol{\omega})$ represents the Jacobian of $\begin{bmatrix}
       \hat{\boldsymbol{u}}(\boldsymbol{\omega})\\
         \boldsymbol{\omega}
    \end{bmatrix}$ with respect to $\boldsymbol{\omega}$ and $\boldsymbol{\Sigma}_{\boldsymbol{\omega}}(\boldsymbol{\omega})$ is the covariance matrix of $\boldsymbol{\omega}$. 
    For the second term we take the zero-th order Taylor expansion around $\boldsymbol{\omega} = \hat{\boldsymbol{\omega}}$, which simplifies the second term to $ \begin{bmatrix}
\boldsymbol{\Sigma}_{\boldsymbol{u}}(\hat{\boldsymbol{\omega}}) & 0 \\
      0 & 0
    \end{bmatrix}$.

Based on the normal approximation to the posterior distribution in (7) of the main text, the approximate $(1-\alpha)$ credible interval for the $k-$th element in $(\boldsymbol{u},\boldsymbol{\omega})^\top$ is given by 
$
    \begin{bsmallmatrix}
        \hat{\boldsymbol{u}}(\hat{\boldsymbol{\omega}})\\
        \hat{\boldsymbol{\omega}}
    \end{bsmallmatrix}_k\pm z_{\alpha/2}\cdot \sqrt{\left(\boldsymbol{J}^\top (\hat{\boldsymbol{\omega}})\hat{\boldsymbol{\Sigma}}_{\boldsymbol{\omega}}(\hat{\boldsymbol{\omega}})\boldsymbol{J}(\hat{\boldsymbol{\omega}})+ \begin{bsmallmatrix}
     \boldsymbol{\Sigma}_{\boldsymbol{u}}(\hat{\boldsymbol{\omega}}) & 0 \\
      0 & 0)
    \end{bsmallmatrix}\right)_{kk}},
$
where   $\boldsymbol{J} (\boldsymbol{\omega})$ represents the Jacobian of $\begin{bsmallmatrix}
       \hat{\boldsymbol{u}}(\boldsymbol{\omega})\\
         \boldsymbol{\omega}
    \end{bsmallmatrix}$ with respect to $\boldsymbol{\omega}$. 

\subsection{Numerical Validation}
This section presents a numerical validation by comparing MAGI-ME with MCMC sampling from the full posterior distribution (which we shall call `MAGI-MCMC') under varying denseness of the discretization set $\boldsymbol{I}$. Additionally, we compare the computational time across these scenarios. To further assess the accuracy of our inference results, we compare the final estimates obtained using a sufficiently dense discretization set with those produced by the \texttt{nlme()} function, where the mean model is pre-specified as the analytical solution to the ODE.

\subsubsection{Benchmark Mixed-effects ODE Model}

As a benchmark model, we consider the population growth model in \citet{wang2014estimating}. This model assumes that the population growth rate in one generation is directly related to the current population size. The population dynamics can be described using the ODE:

\begin{equation}
    x'_j(t)=-\theta_j x_j(t),
    \label{eqn:ode_simple}
\end{equation}
where $x_j(t)$ denotes the population for the $j$-th species, and $\theta_j$ represents the growth rate. We set the true value for fixed-effects parameter $\eta = 3$ and generate the ODE parameter  $\theta_j=\eta+b_j$ for 20 individual species, where $b_j \sim N\left(0, \sigma_b^2\right)$. The initial population $x_j(0)$ is assumed to follow the normal distribution with mean $x_0=1$ and variance $\sigma_0^2=0.1^2$. To create the underlying trajectory, we solve the ODE defined in \eqref{eqn:ode_simple} over the time interval of interest, $[0,1]$, with $|\boldsymbol{\gamma}_j| = 21$ equally-spaced time points. We assume a noisy observation taken at time $t$ follows a Normal distribution with mean $x_j(t)$ and variance $\sigma^2 = 0.1^2$. 

\subsubsection{Comparison between MAGI-ME and MAGI-MCMC}

We vary the denseness of the discretization set $|\boldsymbol{I}_j|= \{21, 41, 81\}$ by inserting 0, 1, 3 discretization points between each pair of adjacent observation time points. We choose $\nu = 2.01$ for the Matern covariance kernel. We set the hyper-parameters and starting values of parameters as described in Section 3.3 of the main text. Diffuse priors are placed on all of the parameters: $\log(\eta)$, $\log(\sigma)$, and $\log(\sigma_b)$ are uniform over all real numbers. The posterior distribution for the MAGI-ME method is implemented in \verb!C++! with optimization performed using the BFGS algorithm. For MAGI-MCMC, the posterior distribution is implemented in $Stan$ \citep{carpenter2017stan} and the default sampling algorithm NUTS (No-U-Turn sampler, \citet{hoffman2014no}) is chosen. We run 5,000 MCMC iterations with the first 2,500 discarded as burn-in. The posterior means from the MCMC samples are taken as the parameter estimates and inferred trajectories. 

To compare these approaches, we compute the parameter RMSEs and empirical coverage probabilities of the 95$\%$ credible interval. To assess the quality of the inferred trajectory, we also compute the MSE between the inferred trajectory $\hat{\boldsymbol{x}}_j(\boldsymbol{\gamma}_j)$  and the truth $\boldsymbol{x}_j(\boldsymbol{\gamma}_j)$ over the observation period, and take the average among the subjects.  \ref{tab:param_mcmc_tmb} summarizes the RMSEs of estimated parameters 
and MSEs of the inferred trajectories. First, we examine the effect of the denseness of the discretization set. Both MAGI-ME and MAGI-MCMC demonstrate a notable decrease in the RMSEs of estimated parameters and MSEs of inferred trajectories, along with empirical coverage probabilities of the 95$\%$ credible interval approaching 0.95, as $|\boldsymbol{I}_j|$ increases from 21 to 41. The inference results with $|\boldsymbol{I}_j| = 41$ can be considered stable; a further increase of $|\boldsymbol{I}_j|$ to 81 only yields a slight improvement for the increased computation cost. 
Second, MAGI-ME and MAGI-MCMC produce comparable parameter estimates and inferred trajectories across all three scenarios with varying discretization levels. MAGI-ME achieves substantial computational gains (which ranges from two to three orders of magnitude faster) over the same MAGI framework using MCMC sampling across varying discretization levels. MAGI-ME's advantage of computational efficiency becomes more pronounced as the discretization level increases.
\begin{table}[hbt!] 
\caption{Table S1: Average parameter estimates (with parameter RMSEs
and empirical coverage probabilities of the 95$\%$ credible interval)  for the mixed-effects model in \eqref{eqn:ode_simple}, comparing MAGI-ME and MAGI-MCMC across 100 simulated datasets, under varying denseness of the discretization set $|\boldsymbol{I}_j|$. The second last column gives the average MSE between the inferred trajectory and the truth, while the last column gives the average runtime (in minutes, on a single CPU core).}
\resizebox{\textwidth}{!}{
\begin{tabular}{lcrrrrrrrrrrr}
\toprule
\multicolumn{1}{c}{\multirow{2}{*}{$|\boldsymbol{I}_j|$}} & \multirow{2}{*}{Method} & \multicolumn{3}{c}{$\eta$} & \multicolumn{3}{c}{$\sigma$} & \multicolumn{3}{c}{$\sigma_b$} & \multicolumn{1}{c}{\multirow{2}{*}{Traj*$10^5$}} & \multicolumn{1}{c}{\multirow{2}{*}{Runtime}} \\ 
\cline{3-11}
\multicolumn{1}{c}{} & & \multicolumn{1}{c}{Est} & \multicolumn{1}{c}{RMSE} & \multicolumn{1}{c}{Cvg} & \multicolumn{1}{c}{Est} & \multicolumn{1}{c}{RMSE} & \multicolumn{1}{c}{Cvg} & \multicolumn{1}{c}{Est} & \multicolumn{1}{c}{RMSE} & \multicolumn{1}{c}{Cvg} & \multicolumn{1}{c}{} & \multicolumn{1}{c}{} \\ 
\hline
\multicolumn{1}{c}{21} & MAGI-ME & 2.95 & 0.082 & 0.86 & 0.030 & 0.0010 & 0.95 & 0.28 & 0.056 & 0.91 & 9.10 &  0.028\\
                       & MAGI    & 2.95 & 0.083 & 0.88 & 0.030 & 0.0010 & 0.95 & 0.29 & 0.054 & 0.93 & 9.17 & 4.655 \\
                       &         &      &       &      &      &        &      &      &       &      &      &       \\
\multicolumn{1}{c}{41} & MAGI-ME & 2.98 & 0.073 & 0.95 & 0.030 & 0.0010 & 0.95 & 0.29 & 0.054 & 0.92 & 8.47 &  0.088 \\
                       & MAGI    & 2.98 & 0.073 & 0.96 & 0.030 & 0.0010 & 0.95 & 0.30 & 0.056 & 0.95 & 8.51 & 48.399 \\
                       &         &      &       &      &      &        &      &      &       &      &      &       \\
\multicolumn{1}{c}{81} & MAGI-ME & 2.99 & 0.071 & 0.95 & 0.030 & 0.0010 & 0.95 & 0.29 & 0.054 & 0.93 & 8.29 &  0.540 \\
                       & MAGI    & 2.99 & 0.071 & 0.97 & 0.030 & 0.0010 & 0.95 & 0.30 & 0.056 & 0.95 & 8.34 &  491.634 \\
\bottomrule
\end{tabular}
}
\label{tab:param_mcmc_tmb}
\end{table}
\subsubsection{Comparison between MAGI-ME and NLME}
If the ODEs have analytic solutions, the mixed effects ODE model can be estimated using the \texttt{nlme()} function in R by defining the mean model as the analytic solution. The analytic solution to the ODE in \eqref{eqn:ode_simple} is given by $$x_{i}(t) = x_{i}(0)\exp(-\theta_i t).$$ 
The \texttt{nlme()} function provides maximum likelihood estimates (MLEs) for fixed-effects parameters and random-effects parameters with corresponding 95$\%$ confidence intervals via the asymptotic normality of the MLE.  
Ideally, methods that approximate the ODE solution should produce parameter estimates comparable to those obtained using the method that pre-specifies the analytic solution. Therefore, we further validate MAGI-ME by comparing its inference results under the most accurate discretization set, i.e., $|\boldsymbol{I}_j|=81$, with the results from the \texttt{nlme()} function. 

For validation of the parameter inference, we use the parameter RMSE and empirical coverage probability of the 95$\%$ credible interval as the performance metrics. The trajectory MSE is also computed for these two methods based on the estimates of the parameters and initial conditions. Recall that the trajectory MSE is obtained as follows: first, we use the numerical solver to construct the ground truth for each subject by solving \eqref{eqn:ode_simple} based on the true parameter values over the given observation period; then, we use the numerical solver to reconstruct the trajectory implied by the estimated parameters from each subject; last, we compute the subject-specific MSE between the true trajectory and the reconstructed trajectory at the observation time points, and take the average among all the subjects.  

The parameter estimates and trajectory MSEs are summarized in \ref{tab:param_tmb_nlme}. This provides further validation of our method, demonstrating that when the discretization set is sufficiently dense, the inference results obtained using MAGI-ME align closely with those from NLME. Specifically, the parameter estimates, RMSEs, and empirical coverage probabilities of the 95$\%$ credible intervals are comparable between the two methods. Additionally, the trajectory MSE values indicate that both methods yield reconstructed trajectories that closely resemble the true trajectories derived from the analytic solution.
\begin{table}[hbt!]
\caption{Table S2: Average parameter estimates obtained by MAGI-ME and NLME (with parameter RMSEs and empirical coverage probability of the 95$\%$ credible interval) for the mixed-effects ODE in \eqref{eqn:ode_simple} across 100 simulated data sets. For MAGI-ME, we use the discretization set $|\boldsymbol{I}_j|=81$ for each subject. }
\resizebox{\textwidth}{!}{
\begin{tabular}{ccrrcrrcrrc}
\toprule
\multirow{2}{*}{Method} & \multicolumn{3}{c}{$\eta$}                                                    & \multicolumn{3}{c}{$\sigma$}                                                   & \multicolumn{3}{c}{$\sigma_b$}                                                & \multirow{2}{*}{MSE*$10^5$} \\ \cline{2-10}
                        & Est                      & \multicolumn{1}{c}{RMSE} & \multicolumn{1}{c}{Cvg} & Est                       & \multicolumn{1}{c}{RMSE} & \multicolumn{1}{c}{Cvg} & Est                      & \multicolumn{1}{c}{RMSE} & \multicolumn{1}{c}{Cvg} &                              \\ \hline
MAGI-ME                 & \multicolumn{1}{r}{2.99} & 0.071                    & 0.95                    & \multicolumn{1}{r}{0.030} & 0.0010                   & 0.95                    & \multicolumn{1}{r}{0.29} & 0.054                    & 0.93                    & \multicolumn{1}{r}{8.30}     \\
NLME                    & \multicolumn{1}{r}{3.00} & 0.071                    & 0.93                    & \multicolumn{1}{r}{0.030} & 0.0010                   & 0.95                    & \multicolumn{1}{r}{0.28} & 0.055                    & 0.91                    & \multicolumn{1}{r}{8.17}     \\ \bottomrule
\end{tabular}
}
\label{tab:param_tmb_nlme}
\end{table}

\section{Method Comparison Details}
In this section, we detail the implementation steps to fit the benchmark mixed-effects ODE model in (8) of the main text. Moreover, we showcase the inferred trajectory along with the 95$\%$ credible interval obtained by MAGI-ME. 
\subsection{Implementation}
We compare MAGI-ME with other representative methods for mixed-effects ODE inference: the numerical solver and collocation-based methods as described in Section 4.1 of the main text. In this section, we provide implementation details
for parameter inference given the simulated data sets. To initialize the numerical solver  and collocation methods, we specify the starting parameter values as follows: we randomly draw starting parameter values of $\eta_1$, $\eta_2$ from the uniform distribution with bound $[0,10]$; the diagonal elements of $\boldsymbol{\Sigma}_{\boldsymbol{b}}$ are randomly drawn from the inverse of a log-normal distribution with mean $-1$ and standard deviation $1$, while the off-diagonal elements are set to zero; the initial condition $x_j(0)$ is set to 0 for each subject and the noise parameter $\sigma$ is initialized at 1. 
For the numerical solver-based method, two approaches were employed for parameter inference, as mentioned in Section 4.1 of the main text: (1) maximizing the joint posterior distribution, referred to as `Numerical-MAP', and (2) drawing samples from the posterior via MCMC, referred to as `Numerical-MCMC'. For Numerical-MAP, we use the $\mathtt{optimizing}()$ function from the \texttt{rstan} R package \citep{carpenter2017stan}, with the BFGS algorithm selected for optimization. As noted by \citet{liang2008parameter}, optimization-based algorithms using the numerical solver may only converge to the local optima due to the sensitivity of the numerical solution to the ODE parameters. 
To mitigate such convergence issues, Numerical-MAP runs 20 tries from different starting parameter values and selects the best parameter set corresponding to the largest log-posterior values. 
For Numerical-MCMC, we adopt the No-U-Turn Sampler (NUTS; \citealp{hoffman2014no}) as the default MCMC algorithm, running 10{,}000 iterations and discarding the first 5{,}000 as burn-in. We use the posterior means obtained from MCMC samples as the parameter estimates. 
For the collocation method implementation of \citet{liu2019bayesian}, we follow the authors’ guidelines when running the code: we place one knot at each observation point and obtain 1000 posterior samples by thinning (i.e., taking every 10th sample) from a total of 10,000 MCMC iterations after burn-in. The parameter estimates are obtained by taking the posterior means of the MCMC samples. 
Although placing denser knots in regions of high-frequency oscillation could potentially improve inference accuracy, the authors' package cannot be easily modified for this purpose. %

For MAGI-ME, we insert 3 additional equally-spaced discretization time point between each pair of adjacent observations, i.e., $\boldsymbol{I}_j = \{0, 0.25, \cdots, 20\}$ for each subject $j=1,\cdots, 25$, which provides stable parameter inference without need for a further increase in $|\boldsymbol{I}_j|$. We take the Matern kernel with $\nu = 2.01$ to accommodate for wiggly system trajectories. To run MAGI-ME, we obtain the starting values for BFGS optimization and estimate the GP hyperparameters as described in Section 3.3 of the main text. 

To ensure a fair comparison, we place the same diffuse priors on the model parameters for each of the methods: $\eta_1,\eta_2,x_j(0)$ are $N(0, 1000)$, $\sigma^2$ is inverse-gamma with shape parameter 0.01 and scale parameter 0.01, and $\boldsymbol{\Sigma}_b$ is inverse-wishart with degree of freedom $p+1$ and scale matrix $\mathbb{I}_p\cdot 0.01$, where $p$ denotes the dimension of the vector $\boldsymbol{\theta}_j$, and $\mathbb{I}_p$ is the identity matrix with dimension $p\times p$. Noting that the collocation method treats the initial condition $x_j(0)$ as part of $\boldsymbol{\theta}_j$, we set $p=4$ for the collocation method, while $p=3$ for the other methods.
\subsection{Inferred Trajectory using MAGI-ME}

\begin{figure}[hbt!]
    \centering
    \includegraphics[width=\textwidth]{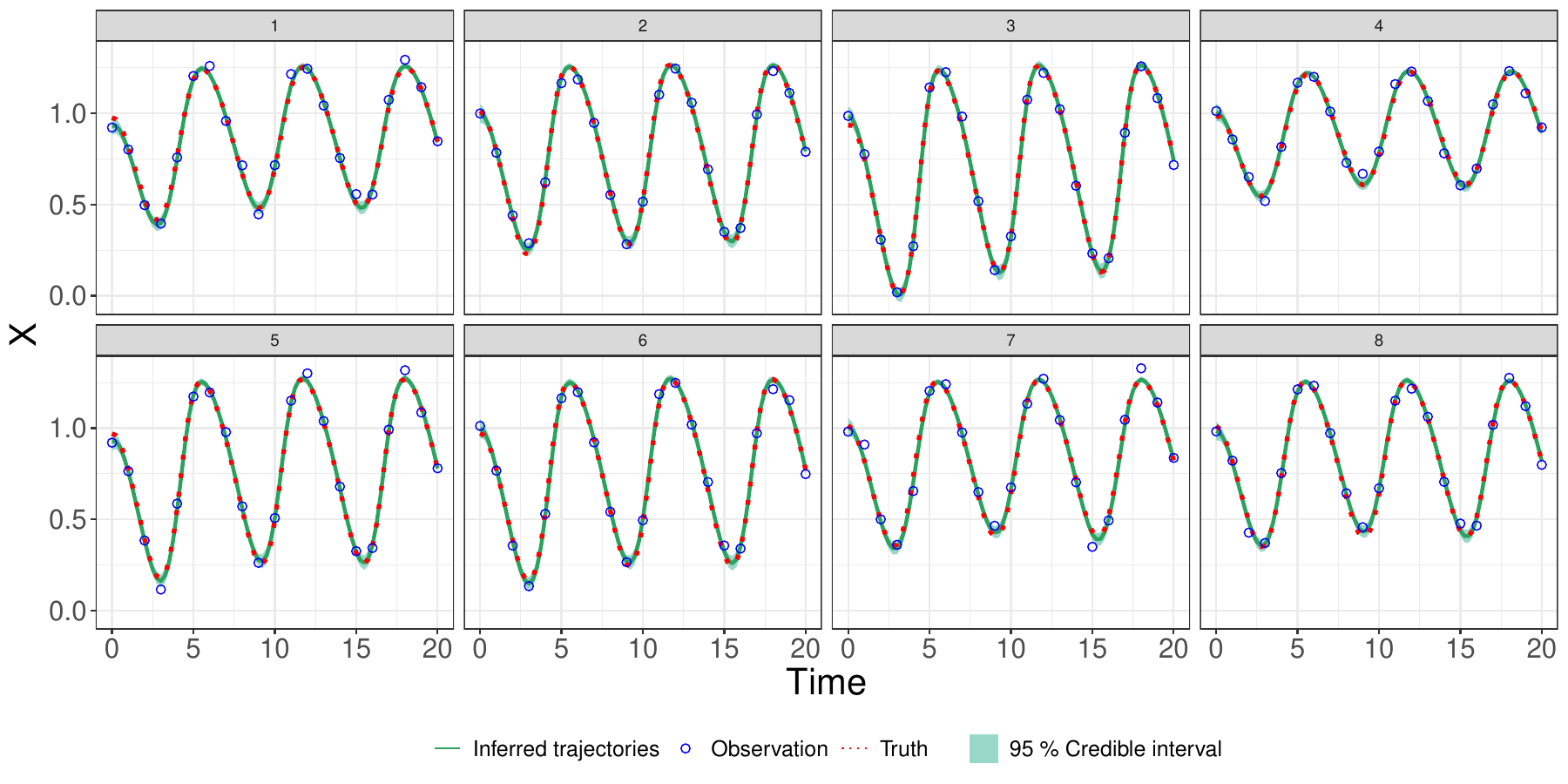}
    \caption{Figure S1: Inferred trajectory of the mixed-effects ODE in (8) of the main text using MAGI-ME. The hollow blue dots are the noisy observations. The green solid line represents the inferred trajectory of the first 8 subjects from the first simulated data set, and the red dotted line represents the truth. The green shaded area is the 95$\%$ pointwise credible interval.}
    \label{fig:traj_simulation_1}
\end{figure}

\ref{fig:traj_simulation_1} presents the inferred trajectories and 95$\%$ credible intervals for the benchmark mixed-effects ODE model using MAGI-ME, which indicate the true trajectory is well-recovered by our method.

\section{FitzHugh–Nagumo Model Details}
\subsection{Implementation}
This section presents the implementation details of MAGI-ME to fit the FitzHugh–Nagumo mixed-effects model as presented in Section 4.2 of the main text. To set up MAGI-ME for this system, we first consider the discretization set. We insert one additional discretization point equally-spaced between each pair of adjacent observations for each subject $j = 1,\cdots, 25$, i.e., $\boldsymbol{I}_j = \{0, 0.25, \cdots, 19.75, 20\}$. Further increasing $|\boldsymbol{I}_j|$ yielded similar inference. Second, we set $\nu = 2.01$ in the Matern kernel to accommodate rougher system trajectories. Third, we place diffuse priors on all the parameters: $\log(\eta_1),\log(\eta_2),\log(\eta_3)$, $\log(\sigma_V)$ and $\log(\sigma_R)$ are uniform over all real numbers; $\boldsymbol{\Sigma}_{\boldsymbol{b}}$ is  inverse wishart with degree of freedom 4 and scale matrix $\mathbb{I}_3\cdot 0.01$. 
\subsection{Inferred Trajectory for FitzHugh–Nagumo Equations using MAGI-ME}
\begin{figure}[hbt!]
    \centering
    \includegraphics[width=\linewidth]{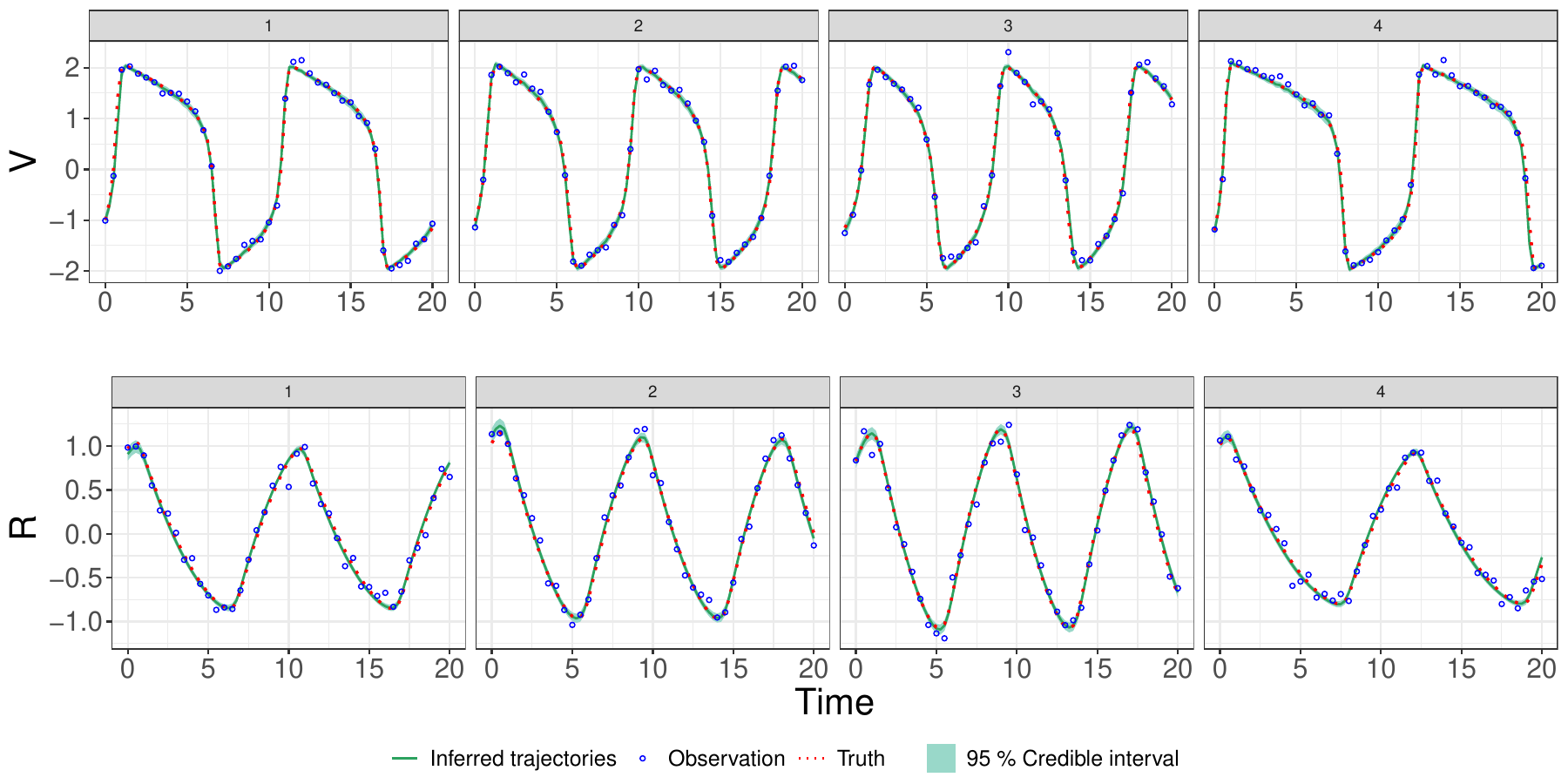}
    \caption{Figure S2: Inferred trajectories of the mixed-effects FN model in (9) of the main text for each component as obtained by MAGI-ME. The green solid line represents the inferred trajectory of the first 4 subjects from the first simulated data set and the red dotted line represents the true trajectory. The green shaded area represents the corresponding 95$\%$ approximate pointwise credible interval. The hollow blue dots are the noisy observations.  }
    \label{fig:traj_plot_2}
\end{figure}

\ref{fig:traj_plot_2} presents the inferred trajectories and 95$\%$ credible intervals for the two-component mixed-effects FitzHugh–Nagumo model using MAGI-ME, which indicate the true trajectory is well-recovered.

\section{Pharmacokinetic Model Details}
In this section, we provide implementation details to fit the pharmacokinetic model using MAGI-ME described in Section 5 of the main text. Then, we present the inferred trajectory together with the 95$\%$ credible interval using the 12-hour IDV plasma concentration data. 
\subsection{Implementation}
 According to \citet{wang2014estimating}, we first set $D_j = 400$ for all subjects receiving treatment I, and $D_j = 600$ for those receiving treatment II. Second, we consider the discretization set. We begin from the smallest evenly spaced set containing all the observation time points $\boldsymbol{I}_{j,0}= \{0, 0.5,\cdots, 12\}$. We construct the discretization set $\boldsymbol{I}_j = \{0, 0.125, \cdots, 12\}$ for all subjects in the two treatment groups and find that further increasing $|\boldsymbol{I}_j|$ yields similar inference results. Third, we use the Matern kernel with $\nu = 2.01$ as the default choice. 
We select the GP bandwidth hyperparameter to minimize the sum of squared errors (SSE) between the observations and the inferred trajectories over the fitting period for each subject in the two treatment groups. Fourth, we choose the priors for the fixed-effects parameters, noise-level parameters and random-effects parameters. To ensure the fixed-effects parameters remain meaningful with positive values, we reparametrize them on a logarithmic scale and adopt the same prior as suggested by \citet{liu2019bayesian}. Specifically, we set a multivariate normal prior $N\left((-0.30, -1.0, -3.0)^\top,\mathbb{I}_3*1000\right)$ for $(\log(Ka), \log(Ke), \log(Cl))^\top$. Diffuse priors are placed on the variance parameters: $\log(\sigma)$, $\log(\sigma_{Cl})$, and $\log(\sigma_{Ka})$ are uniform over all real numbers. Last, we use the BFGS algorithm to perform optimization. 
\subsection{Inferred Trajectory for Pharmacokinetic Model}
\begin{figure}[htbp]
    \centering
    \begin{subfigure}{\linewidth}
        \centering
        \includegraphics[width=0.9 \linewidth]{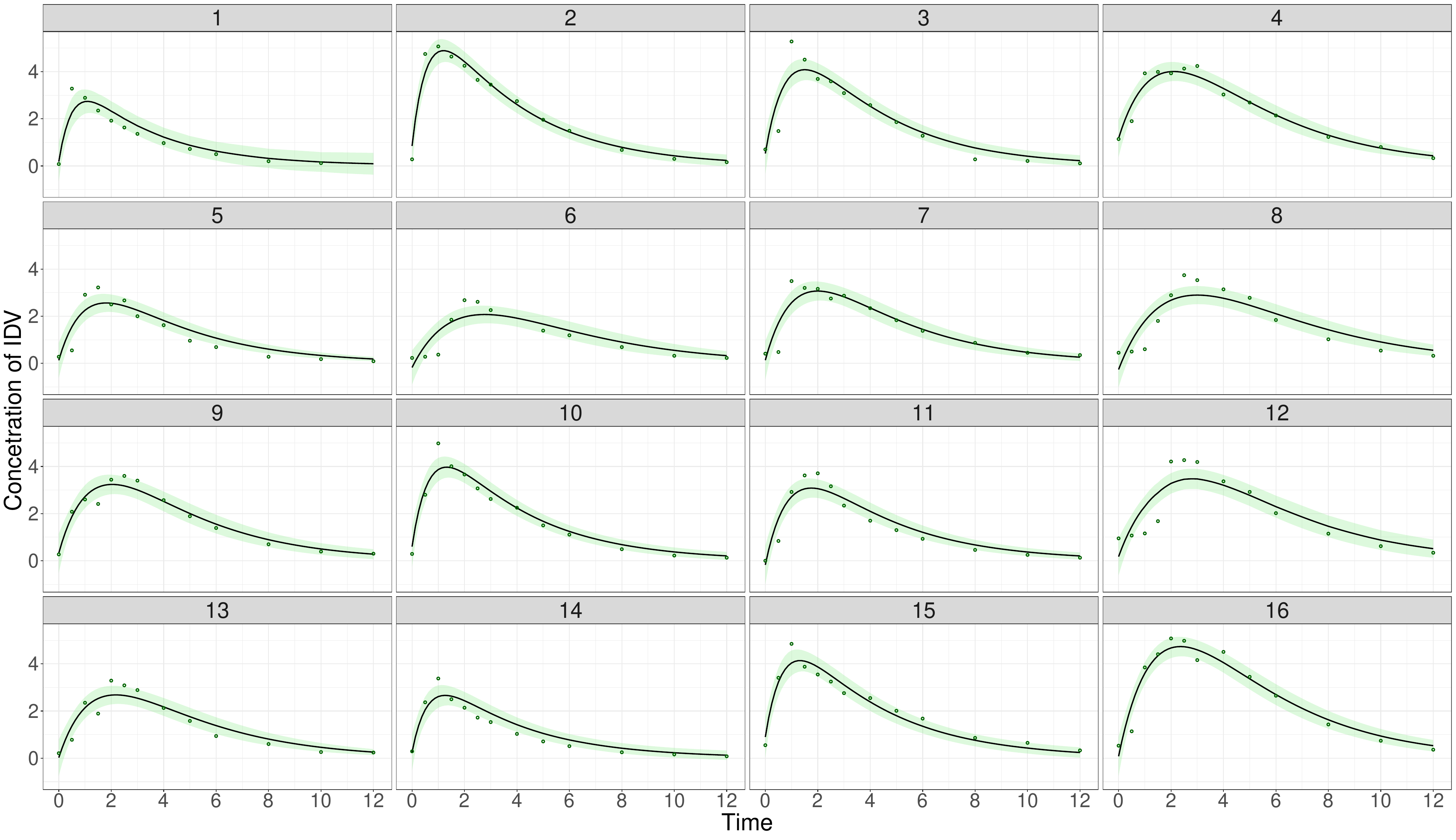}
        \caption*{(a) Inferred trajectories for subjects receiving Treatment I.}
    \end{subfigure}

    \begin{subfigure}{\linewidth}
        \centering
        \includegraphics[width=0.9 \linewidth]{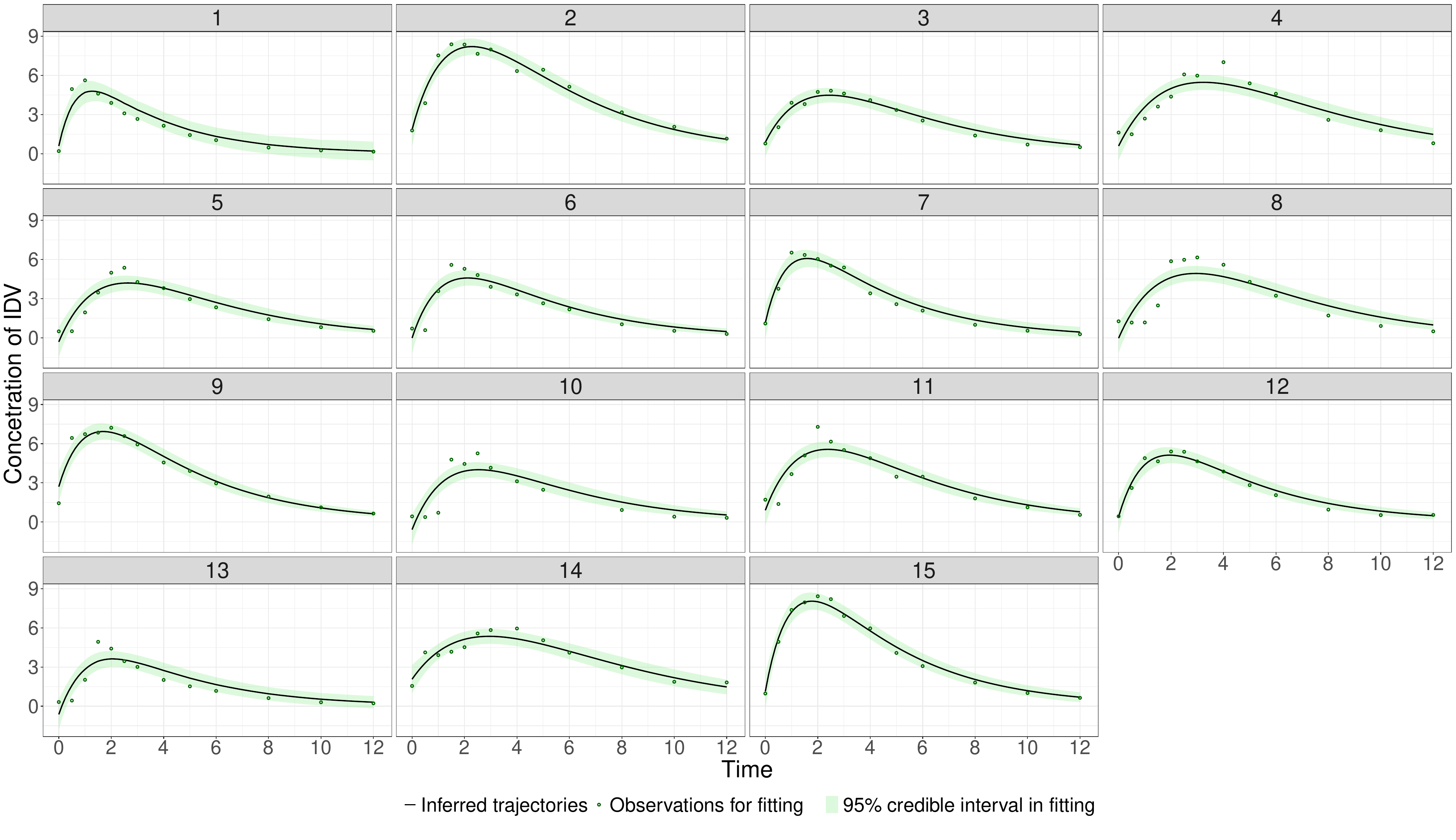}
        \caption*{(b) Inferred trajectories for subjects receiving Treatment II.}
    \end{subfigure}
    
    \caption{Figure S3: Inferred trajectories for the pharmacokinetic mixed-effects model, using IDV concentration measurements  over the 12-hour period. The black line represents the inferred trajectory and the green shaded area represents the 95$\%$ pointwise credible interval. The green dots are the observed data. The top panel corresponds to the subjects taking Treatment I and the bottom panel corresponds to those taking Treatment II. }
    \label{fig:full_data_comparison}
\end{figure}

\ref{fig:full_data_comparison} presents the inferred trajectories and 95$\%$ credible intervals of IDV concentration for each subject in the real data using MAGI-ME.  
\end{document}